\documentclass[aps,preprint]{revtex4}%
\usepackage{amsfonts}
\usepackage{amsmath}
\usepackage{amssymb}
\usepackage{graphicx}
\usepackage{xcolor}
\usepackage{hyperref}
\usepackage{lineno,hyperref}
\hypersetup{colorlinks =true, allcolors = blue}
\providecommand{\U}[1]{\protect\rule{.1in}{.1in}}

\begin{document}
\title{Modified Cosmic Chaplygin AdS Black Hole}
\author{Ujjal Debnath}\email{ujjaldebnath@gmail.com}
\affiliation{Department of Mathematics, Indian Institute of Engineering Science and Technology, Shibpur, Howrah-711 103, India.}
\author{Behnam Pourhassan}\email{b.pourhassan@du.ac.ir}\email{b.pourhassan@candqrc.ca}
\affiliation{School of Physics, Damghan University, Damghan, 3671641167, Iran.}
\affiliation{Canadian Quantum Research Center 204-3002 32 Ave Vernon, BC V1T 2L7 Canada.}
\author{\.{I}zzet Sakall{\i}}\email{izzet.sakalli@emu.edu.tr}
\affiliation{Physics Department, Eastern Mediterranean
University, Famagusta 99628, North Cyprus via Mersin 10, Turkey.}

\begin{abstract}
In this paper, we construct a new charged AdS black hole with modified cosmic Chaplygin gas (MCCG). In comparison to the previous model (modified Chaplygin AdS black hole) existed in the literature, we now have two new parameters: the cosmic parameter and the black hole's electric charge. We examine the asymptotically charged AdS black hole thermodynamics with MCCG under the assumption of a negative cosmological constant as a thermodynamics pressure. Using MCCG, we developed a new solution to Einstein's AdS black hole field equations.The mass parameter, volume, electric potential, entropy, and temperature of the charged AdS black hole have all been computed. We also study the various energy conditions for the MCCG as a source fluid of AdS black hole. For some restrictions on the involved parameters, we show that these conditions are met. Then, we study the thermodynamical stability, critical points, and Joule-Thompson expansion for the back hole obtained. We reveal that while the existence of cosmic parameter yields to a  complete stable mode, its absence gives rise to some unstable regions. Then, using MCCG, we treat the obtained black hole's thermodynamics as a heat engine and calculate the work done as well as the heat engine's maximum efficiency in the Carnot cycle. The work done, the efficiency, and the Rankine cycle's efficiency are also investigated. All of the findings are graphically depicted and discussed.

\end{abstract}

\volumeyear{ }
\eid{ }
\date{\today}
\received{}

{
\let\clearpage\relax
\maketitle
}

\section{Introduction} \label{sec1}
Black hole thermodynamics \cite{01} is one of the important fields of research in theoretical physics. It helps to obtain knowledge about quantum theory of gravity. It has been found that the black hole entropy (hence black hole thermodynamics) depends on the event horizon area \cite{02}, except for 2-dimensional black holes \cite{2D1} where entropy function formalism is used to calculate the black hole entropy depends on the dilaton field at the horizon \cite{2D2, 2D3, 2D4, 2D5}. There are several kinds of black holes which have different geometries and hence different thermodynamics. The Schwarzschild black hole is the simplest one where it is found that the black hole temperature decreases by absorbing mass \cite{03}. The thermodynamics of a Reissner-Norstr\"{o}m (RN) black hole \cite{RN} is also found to be identical to that of a standard black hole \cite{04, 05}. Another interesting kind of black holes is the Ho\^{r}ava Lifshitz black holes which have been studied thermodynamically \cite{HL1, HL2, HL3, HL4}. Black holes with hyperscaling violation takes considerable attentions and being used to study the condensed matter physics \cite{hyper1, hyper2, hyper3,Gursel:2020xou,Gursel:2019fyd}. G\"{o}del-type universes are yet another intriguing spacetime that allows for time travel, and they are already being studied in the fields of thermodynamics and statistics \cite{G1, G2, G3}. Similar works were also done for Myers-Perry \cite{MP1, MP2, MP3} and STU black holes \cite{STU1, STU2, STU3}. Hairy black holes, which can be thought of as bound states of conventional black holes without hair, are fascinating type of black holes from a variety of perspectives \cite{h1, h2, h3, h4, h5, h6, h7}. Anti-de Sitter (AdS) black holes \cite{Gibb, AdS1, AdS2} are also interesting kind of compact objects that are widely employed in the holographic works \cite{ho1, ho2, ho3}. The thermodynamics of the Schwarzschild-AdS black hole was explored by Ref. \cite{Hawking}. In the sequel, the thermodynamics of non-rotating charged RN-AdS black hole was investigated by the Refs. \cite{Cham1,Cham2}. By analyzing PV criticality \cite{PV1, Del, PV2, PV3}, it is interesting to see that some black holes are holographic dual of van der Walls fluids \cite{van1, van2, van3}. For example, charged rotating AdS black holes behave like a van der Walls fluid \cite{Cv, Niu, rostam}. Obtaining van der Walls behavior helps us to obtain new perspective on black hole thermodynamics, like identifying the black hole mass with chemical enthalpy \cite{Kubi0}. All above works motivate the physicists to study more and more on the black hole thermodynamics, specially the charged AdS black holes, where the cosmological constant may be considered as thermodynamic
pressure \cite{Del, van1, Kubi, Gun, Cald, Cre, Belt},
\begin{equation}\label{000}
p=-\frac{\Lambda}{8\pi},
\end{equation}
in which
\begin{equation}
\Lambda=-\frac{(d-1)(d-2)}{2l^{2}}.
\end{equation}
$\Lambda$ denotes the (negative) cosmological constant in $d$-dimensional spacetime. Throughout this study, we shall consider the 4-dimensional AdS spacetime, thus
$\Lambda=-\frac{3}{l^{2}}$  by which $l$ is the curvature constant of AdS space. In the case the uncharged non-rotating AdS black hole, the first law of
black hole thermodynamics is given by \cite{Del},
\begin{equation}
dE=TdS-pdV+\Phi dQ,
\end{equation}
where the internal energy $E$ might be considered as the black hole mass \cite{Del}, also $S$, $T$, and $Q$ are the entropy, temperature and electric charge of the black hole, respectively. Moreover, $\Phi$ denotes the chemical potential of the dual fluid. According to the equation (\ref{000}) the negative cosmological constant is considered as a
thermodynamic pressure in the first law of black hole thermodynamics, it gives us motivation to find an asymptotically
charged AdS metric whose thermodynamics matches exactly that of the Chaplygin gas. Among of several kinds of dark energy models (see for example \cite{JHAP}) we are interested to consider Chaplygin gas equation of state. More than a century ago, S. Chaplygin introduced an equation of state in aerodynamics which is given by \cite{Chaplygin},
\begin{equation}\label{0}
p=-\frac{B}{\rho}.
\end{equation}
Today, it is argued that the equation of state (\ref{0}), which is reproduced by string theory \cite{string}, could explain the negative pressure of dark energy \cite{ch1, ch2}. Then, in order to have better explanation of dark energy, generalized Chaplygin gas (GCG) equation of state is constructed \cite{ch3}. Several cosmological models \cite{G1, G2, G3} indicated that GCG is not good model to verify recent observations, hence viscous GCG \cite{vi1, vi2, vi3, vi4, vi5, vi6, vi7} or cosmic GCG \cite{co1, co2} models constructed. In order to have more agreement with observational data, modified Chaplygin gas (MCG) equation of state was introduced in Refs. \cite{M1, M2, M3}. In a recent work, AdS black hole thermodynamics with the MCG has been studied \cite{Debnath1} and, moreover, the effects of viscosity on the MCG are analyzed \cite{vim1, vim2, vim3, vim4}. Also, MCCG was considered in the following Refs. \cite{com1, com2, com3}. In Ref. \cite{com3}, thermodynamics of the MCCG was investigated to analyze the nature of Universe. Since MCCG is a very well-known candidate of dark energy model, so our main motivation is to construct AdS black hole in presence of MCCG and study the thermodynamic nature with heat engine phenomena of such type of black hole.

In this paper, we would like to consider charged AdS black hole with MCCG and study its thermodynamical features. The metric to be obtained is different than the one seen in Ref. \cite{com3} where pure MCCG was considered. Our study is indeed the extension of the Ref. \cite{Debnath1} by including cosmic parameter and electric charge, which is corresponding to switch on chemical potential in the holographic dual picture. We should note that the MCCG is introduced to explain the late-time cosmic acceleration with the aid of a particular fluid, which is a candidate for the dark energy. Our work is indeed a way to obtain more information about such a fluid. We use holographic principles and study black hole thermodynamics which is a dual of MCCG fluid. Moreover, AdS black hole with MCG \cite{Debnath1} has some unstable regions, which are shown that they can be removed with the presence of cosmic parameter.\\
In the next section, we construct MCCG AdS black hole and in the sequel, in section \ref{sec3}, we discuss the required energy conditions and study the horizon structure of the AdS black hole with MCCG. In section \ref{sec4}, we study black hole thermodynamics and discuss the stability and phase transition. In the section \ref{sec5}, we consider the MCCG AdS black hole as heat engine. In section \ref{sec6}, we draw our conclusions and make some suggestions for the future works.

\section{AdS Black Hole with MCCG} \label{sec2}
First of all, in this section, we review how the MCCG equation of state grows up. The MCG equation of state given by \cite{Bena, Debnath},
\begin{equation}
p=A\rho-B\rho^{-\alpha},
\end{equation}
where $A$, $B$, and $\alpha$ are constants. If one sets $A=0$ then MCG is obtained, and by setting $A=0$ together with $\alpha=1$ one recovers the Chaplying gas equation of state. These constant could be fixed by observational data.\\
In 2003, P. F.
Gonz$\acute{a}$lez-Diaz \cite{Gon} introduced the generalized
cosmic Chaplygin gas equation of state. The advantage of this model is stability even if the vacuum fluid satisfies the energy condition of the phantom field. The corresponding EoS is given by
\begin{equation}
p=-{\rho}^{-\alpha}\left[ C + (\rho^{1+\alpha}-C)^{-w}\right]
\end{equation}
where $C=\frac{B}{1+w}-1$, and $B$ is a constant. Also, $-\ell<w\leq0$, where $\ell$ is a positive constant.\\
We are interested to consider more general case. The equation of state for modified cosmic Chaplygin gas (MCCG) is
given by \cite{Pour},
\begin{equation}\label{3}
p=A\rho-\rho^{-\alpha}\left(C+(\rho^{1+\alpha}-C)^{-w} \right).
\end{equation}
Setting $w=0$ yields to MCG equation of state. Refs. \cite{Pour, Naji} considered MCCG from cosmological point of view. The MCCG EoS reduces to that of current Chaplygin unified models for dark matter and dark energy in the limit  $w\rightarrow 0$ and satisfies the following  conditions:\\
(i) it reduces to de Sitter fluid at late time when $w= - 1$,\\
(ii) it reduces to linear EoS $p=w \rho$ when $A\rightarrow 0$,\\
(iii) it reduces to the EOS of Chaplygin unified dark matter models when energy density is very high.\\
This fluid propagates between  dust era in the past and $\Lambda$CDM in the future. We shall construct an asymptotically AdS black hole with MCCG whose
thermodynamics coincides with the above equation of state
(\ref{3}). Now, we assume the static spherically symmetric ansatz as in the following form \cite{Del, van1},
\begin{equation}\label{4}
ds^{2}=-fdt^{2}+\frac{dr^{2}}{f}+r^{2}d\Omega_{2}^{2},
\end{equation}
where
\begin{equation}\label{5}
f\equiv f(r,\rho)=\frac{r^{2}}{l^{2}}-\frac{2M}{r}+\frac{Q^{2}}{r^{2}}-g(r,\rho),
\end{equation}
where $M$ and $Q$ related to the mass and charge respectively, also, the unknown function $g(r,\rho)$ determined so that above ansatz behave like a charged AdS black hole holographic dual of MCCG.\\
Now, we consider the AdS spacetime with the negative cosmological constant $\Lambda$ (vacuum
pressure) from Einstein's equations $G_{\mu\nu}+\Lambda
g_{\mu\nu}=8\pi T_{\mu\nu}$. The entropy, mass parameter, volume, and
temperature of the AdS black hole can be written in terms of (event)
horizon radius $r_{h}$ as in the following forms \cite{Del, van1},
\begin{equation}\label{6}
S=\pi r_{h}^{2},
\end{equation}
which is obtained using area formula,
\begin{equation}\label{7}
M=\frac{4\pi}{3}~r_{h}^{3}p+\frac{Q^{2}}{2r_{h}}-\frac{1}{2}~r_{h}g(r_{h},\rho)~,
\end{equation}
which is obtained using relation (\ref{5}), then the black hole volume is given by
\begin{equation}\label{8}
V=\frac{\partial M}{\partial
p}=\frac{4\pi}{3}~r_{h}^{3}-\frac{1}{2}~r_{h}\frac{\partial
g(r_{h},\rho)}{\partial\rho}(\frac{dp}{d\rho})^{-1},
\end{equation}
and
\begin{equation}\label{9}
T=\frac{1}{4\pi}~\left[\frac{\partial f(r,\rho)}{\partial
r}\right]_{r=r_{h}}.
\end{equation}
The electrostatic potential is obtained as,
\begin{equation}\label{Phi}
\Phi=\frac{\partial M}{\partial
Q}=\frac{Q}{r_{h}}.
\end{equation}
By using the integrability conditions, the first law of thermodynamics becomes \cite{Debnath1},
\begin{equation}\label{10}
S=\frac{\rho+p}{T}~V-\frac{\Phi Q}{T}.
\end{equation}
Now, we can put (\ref{6}), (\ref{7}), (\ref{8}), (\ref{9}) and (\ref{Phi}) in the equation
(\ref{10}) to obtain,
\begin{equation}\label{11}
\left[\frac{9 r_{h}^{2}}{l^{2}}+9\frac{Q^{2}}{r_{h}^{2}}-3g(r_{h},\rho)-3r_{h}\left[\frac{\partial g(r,\rho)}{\partial
r}\right]_{r=r_{h}}\right]\frac{dp}{d\rho}-(\rho+p)\left[16\pi r_{h}^{2}\frac{dp}{d\rho}-6\frac{\partial
g(r_{h},\rho)}{\partial \rho}\right] =0.
\end{equation}
because $p$ is dependent on $\rho$, we can rewrite the above equation as follows,
\begin{equation}\label{11-1}
\frac{9 r_{h}^{2}}{l^{2}}+\frac{9Q^{2}}{r_{h}^{2}}-3g(r_{h}, p)-3r_{h}\left[\frac{\partial g(r,\rho)}{\partial
r}\right]_{r=r_{h}}-(\rho+p)\left[16\pi r_{h}^{2}-6\frac{\partial
g(r_{h}, p)}{\partial p}\right] =0.
\end{equation}
Now, we would like to obtain the unknown function $g(r, p)$, then by using the equations (\ref{3}) to obtain the function of $g(r, \rho)$. To this end, we may
assume a following form,
\begin{equation}\label{12}
g(r, p)=X(r)Y(p),
\end{equation}
where $X(r)$ and $Y(p)$ are arbitrary functions of $r$ and $p$, respectively.
Now, substituting the expression (\ref{12}) in the
equation (\ref{11-1}), we obtain the following solutions of $X$ and $Y$,
\begin{equation}
X(r)=C_{X}e^{\frac{X_{0}r}{3r_{h}}},
\end{equation}
and
\begin{equation}
Y(r)=C_Ye^{\frac{8\pi l^{2}(X_{0}+3)r_{h}^{4}}{27(Q^{2}l^{2}+r_{h}^{4})}p},
\end{equation}
where $X_{0}$ is arbitrary constant, while $C_X$ and $C_Y$ are integration constants.
Substituting the above solutions in equation (\ref{12}) and in the sequel using the equations (\ref{3}), we get the
expression of $g(r,\rho)$ as follows
\begin{equation}\label{17}
g(r,\rho)=C_{XY}\exp{\left(\frac{X_{0}r}{3r_{h}}+\frac{8\pi l^{2}(X_{0}+3)r_{h}^{4}}{27(Q^{2}l^{2}+r_{h}^{4})}\left(A\rho-\rho^{-\alpha}\left(C+(\rho^{1+\alpha}-C)^{-w} \right)\right)\right)},
\end{equation}
where $C_{XY}\equiv C_XC_Y$.

\begin{figure}[h!]
 \begin{center}$
 \begin{array}{cccc}
\includegraphics[width=50 mm]{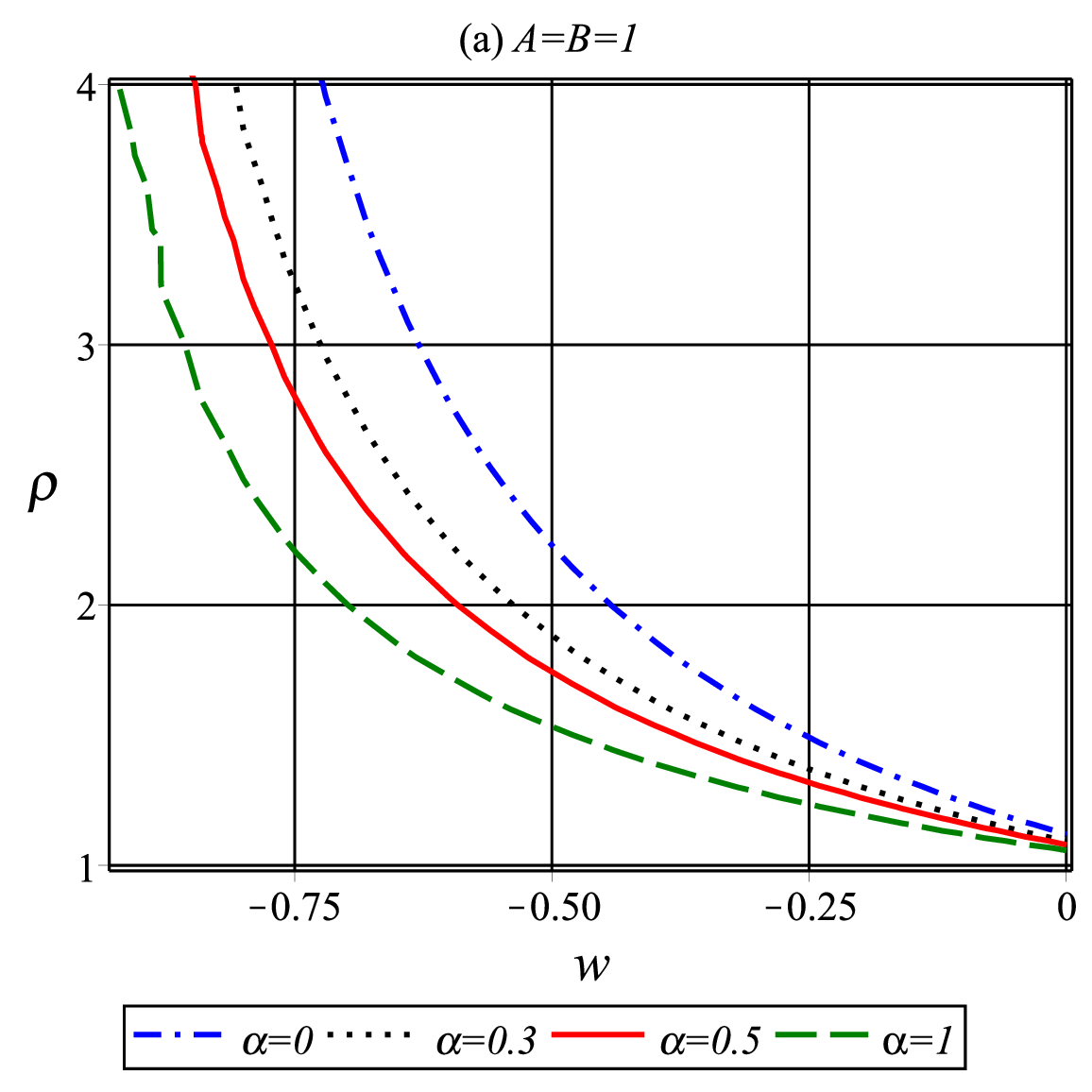}\includegraphics[width=50 mm]{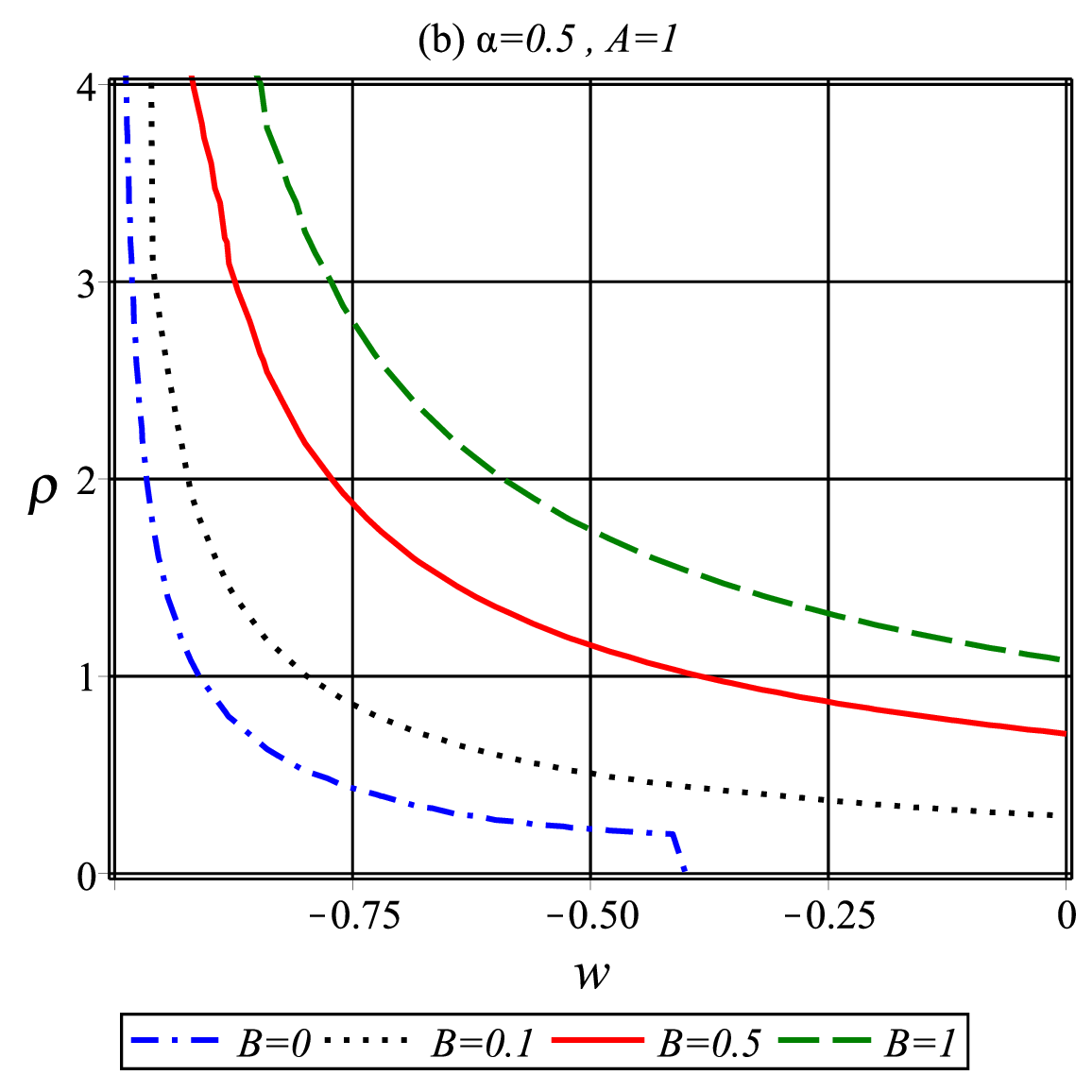}\includegraphics[width=50 mm]{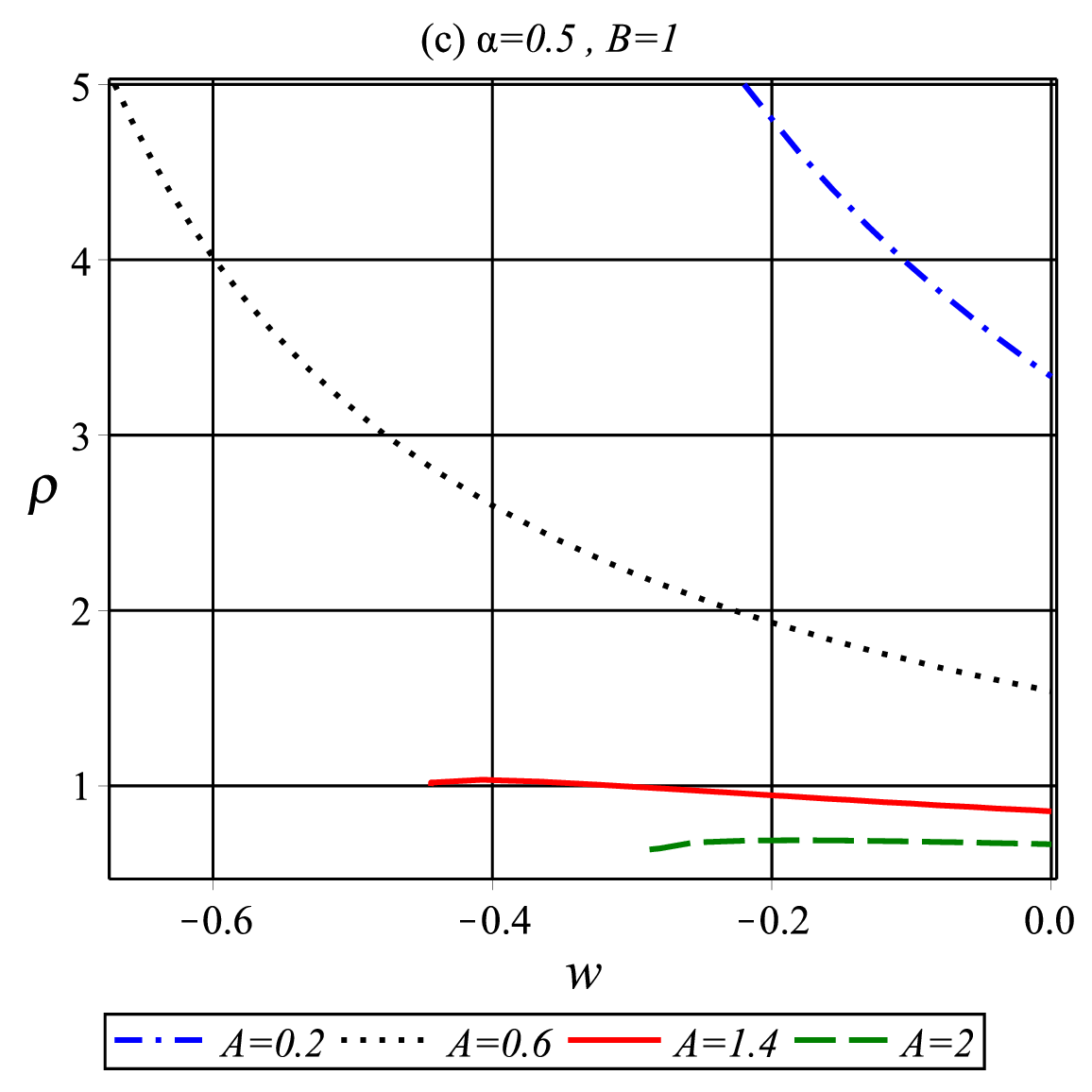}
 \end{array}$
 \end{center}
\caption{$\rho$ in terms of $w$ for $l=1$.}
 \label{fig1}
\end{figure}

Now, putting the solution of $g(r,\rho)$ in equation (\ref{5}), one can
obtain the solution of the function $f(r,\rho)$ as
\begin{equation}\label{18}
f(r,\rho)=\frac{r^{2}}{l^{2}}-\frac{2M}{r}+\frac{Q^{2}}{r^{2}}-C_{XY}\exp{\left(\frac{X_{0}r}{3r_{h}}+\frac{8\pi l^{2}(X_{0}+3)r_{h}^{4}}{27(Q^{2}l^{2}+r_{h}^{4})}\left(A\rho-\rho^{-\alpha}\left(C+(\rho^{1+\alpha}-C)^{-w} \right)\right)\right)}.
\end{equation}
This new form of black hole solution may be called {\it MCCG AdS black hole} (after the names of Van der Waals
black hole \cite{van1} and
MCG black hole \cite{Debnath1}). So, recalling equations (\ref{3}) and
(\ref{18}), we have
\begin{equation}\label{19}
f=\frac{r^{2}}{l^{2}}-\frac{2M}{r}+\frac{Q^{2}}{r^{2}}-C_{XY}\exp{\left(\frac{X_{0}r}{3r_{h}}+\frac{8\pi l^{2}(X_{0}+3)r_{h}^{4}}{27(Q^{2}l^{2}+r_{h}^{4})}p\right)}
\end{equation}
Assuming $X_{0}=r_{h}$ and $C_{f}\equiv e^{\frac{(X_{0}+3)r_{h}^{4}}{9(Q^{2}l^{2}+r_{h}^{4})}p}$ one can rewrite the metric function (\ref{19}) as following,
\begin{equation}\label{19-1}
f=\frac{r^{2}}{l^{2}}-\frac{2M}{r}+\frac{Q^{2}}{r^{2}}-Be^{\frac{r}{3}},
\end{equation}
where $B\equiv C_{XY}C_{f}$ is a constant. It is worth noting that the Schwarzschild limit of the black hole is obtained by by setting $Q=0$ and taking the limit of $l\rightarrow\infty$ which means $r\ll l$; then the first and last terms in r.h.s. of the equation (\ref{19-1}) are negligible comparison with the other terms.\\
To calculate $\rho$, one can use the equation (\ref{3}):
\begin{equation}
(\rho^{1+\alpha}-C)^{w}((A\rho-p)\rho^{\alpha}-C)=1,
\end{equation}
where $p=\frac{3}{8\pi l^{2}}$. We find that in order to have real positive density, the value of cosmic parameter should be in the range $-1<w\leq0$. However, values of $\alpha$, $A$, and $B$ are also important. In the Fig. \ref{fig1}, we plot the density versus cosmic parameter graph for various values of the cosmic parameters. In the Fig. \ref{fig1} (a), one can see the effect of $\alpha$ on the plots. Dashed green line corresponds to generalized cosmic Chaplygin AdS black hole. For the case of $w=0$, there is no differences between various values of $\alpha$. In Fig. \ref{fig1} (b) we vary $B$ parameter. In the case of $B=0$, we have a polytropic fluid where the effect of cosmic parameter is illustrated. In that case, we can see that upper limit of $w$ reduced (see the dash-dotted blue line of the Fig. \ref{fig1} (b)). In the Fig. \ref{fig1} (c) we vary $A$ parameter and find that increasing $A$ reduces the value of $\rho$. In addition, increasing $A$ limits the lower bound of $w$.\\
In the next section, we shall attempt to constraint the model parameters by using the energy conditions. Then, we study the horizon structure of MCCG AdS black hole.

\section{ Energy Conditions} \label{sec3}
In this section, we are going to examine whether the weak, strong, and dominant
energy conditions are satisfied/violated for the source fluid. The
energy-momentum tensor for the anisotropic source fluid is given by \cite{Del, van1},
\begin{equation}\label{fiducial}
T^{\mu\nu}=\left(
\begin{array}{cccc}
\varrho & 0 & 0 & 0 \\
0 & p_{1} & 0 & 0 \\
0 & 0 & p_{2} & 0 \\
0 & 0 & 0 & p_{2}%
\end{array}%
\right),
\end{equation}
where $\varrho$ is the density of energy, while $p_{i}$ ($i=1,2,3$) are the
 source fluid pressures, which are obtained as the following equations \cite{Del, van1},
\begin{eqnarray}
\varrho&=&-p_{1}=\frac{1-f-rf'}{8\pi r^{2}}+p\nonumber\\
&=&\frac{1}{8\pi
r^{2}}\left[1-\frac{3r^{2}}{l^{2}}+\frac{Q^{2}}{r^{2}}+(1+\frac{r}{3})Be^{\frac{r}{3}}\right]+\frac{3}{8\pi l^{2}},
\end{eqnarray}
and
\begin{eqnarray}
p_{2}=p_{3}&=&\frac{rf''+2f'}{16\pi r}-p\nonumber\\
&=&\frac{1}{8\pi r^{2}}\left[\frac{3r^{2}}{l^{2}}+\frac{Q^{2}}{r^{2}}-\frac{r}{6}(2+\frac{r}{3})
Be^{\frac{r}{3}}\right]-\frac{3}{8\pi l^{2}},
\end{eqnarray}
where $G=1$ assumed and prime denotes derivative with respect to $r$.
From the above equations we get the following energy conditions:\\
(i) Weak energy condition requires that both $\varrho\ge 0$ and $\varrho+p_{i}\ge 0$
$(i=1,2,3)$ should be satisfied in overlap region of the following relations:
\begin{eqnarray}
W_{1}\equiv1-\frac{3r^2}{l^2}+\frac{Q^2}{r^2}+(1+\frac{r}{3})Be^{\frac{r}{3}}&\geq&0,\nonumber\\
W_{2}\equiv1+\frac{2Q^2}{r^2}+(1+\frac{r}{3}-\frac{r^2}{18})Be^{\frac{r}{3}}&\geq0.
\end{eqnarray}

\begin{figure}[h!]
 \begin{center}$
 \begin{array}{cccc}
\includegraphics[width=60 mm]{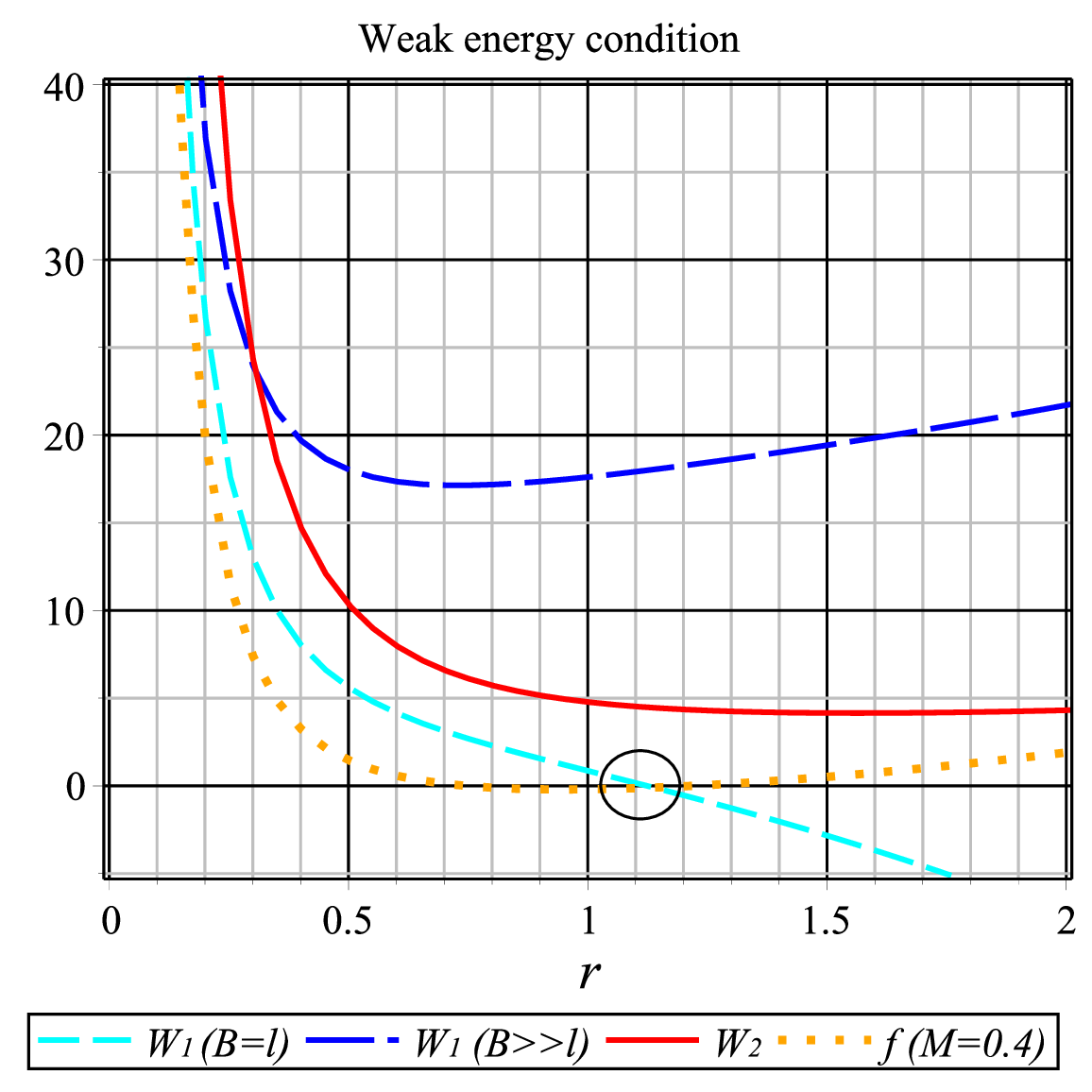}
 \end{array}$
 \end{center}
\caption{Weak energy condition for $Q=1$ and $l=1$.}
 \label{figW}
\end{figure}

In the case of $B\gg l$ both conditions are satisfied in all space around horizon ($r\approx r_{h}$), while in the case of $B\approx l$ we can find that weak energy condition satisfied for $r\leq r_{h}$. In the Fig. \ref{figW} we show the behaviors of $W_{1}$ and $W_{2}$ comparing with the metric function $f$.  Zeroth value of the metric function (horizon radius which will discussed in details later) is represented in circle. The first weak condition for $B\gg l$ ($B=10l$) selected in the long dash blue line of the Fig. \ref{figW}) satisfied in all space while the second condition is not strongly dependent on $B$, and it is positive for $r\approx r_{h}$ (see solid red line of the Fig. \ref{figW}). It will be negative at $r\gg r_{h}$. The first weak condition is also violated for $B=l$ if $r> r_{h}$. Hence, for the region $r\leq r_{h}$ both weak energy conditions are satisfied.

\begin{figure}[h!]
 \begin{center}$
 \begin{array}{cccc}
\includegraphics[width=60 mm]{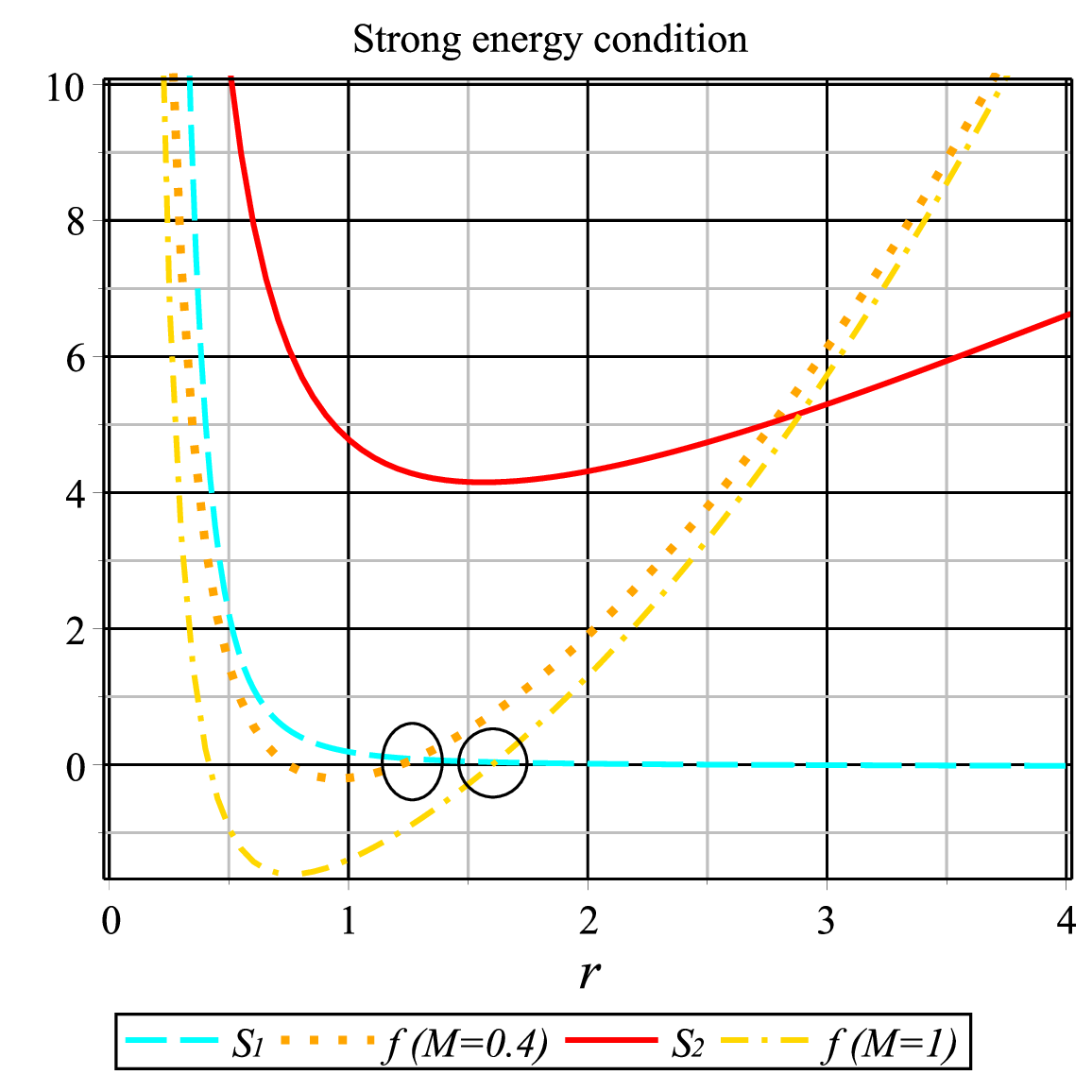}
 \end{array}$
 \end{center}
\caption{Strong energy condition for $Q=1$, $l=1$, and $B=1$.}
 \label{figS}
\end{figure}

(ii) Strong energy condition requires that both $\varrho+\sum_{i}p_{i}\ge 0$ and $\varrho+p_{i}\ge 0$ $(i=1,2,3)$, which should be satisfied in overlap region with the following relations:
\begin{eqnarray}
S_{1}&\equiv&\frac{1}{8\pi r^{2}}\left[1+\frac{3r^2}{l^2}+\frac{3Q^2}{r^2}+(1-\frac{r}{3}-\frac{r^{2}}{9})Be^{\frac{r}{3}}\right]-\frac{3}{8\pi l^{2}}\geq0,\nonumber\\
S_{2}&\equiv&1+\frac{2Q^2}{r^2}+(1+\frac{r}{3}-\frac{r^2}{18})Be^{\frac{r}{3}}\geq0.
\end{eqnarray}
It is clear that the second condition is like previous case and, as before, for the case of $r\leq r_{h}$ the strong energy conditions are also satisfied. The first condition is satisfied for $r\approx r_{h}$ (see dashed cyan line of Fig. \ref{figS}). Zeros of metric function for two different values of mass parameter are represented in circles.

\begin{figure}[h!]
 \begin{center}$
 \begin{array}{cccc}
\includegraphics[width=60 mm]{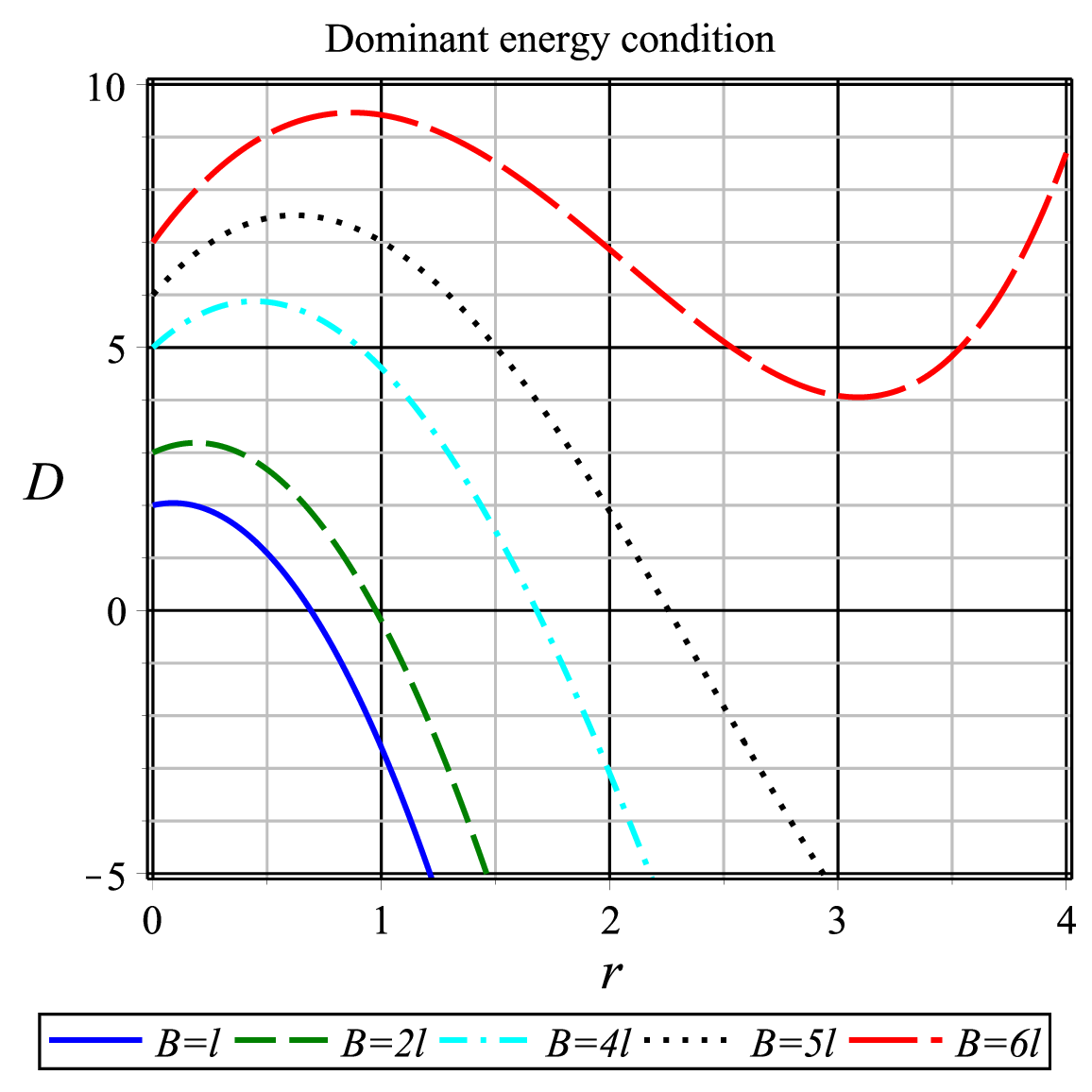}
 \end{array}$
 \end{center}
\caption{Dominant energy condition for $Q=1$ and $l=1$.}
 \label{figD}
\end{figure}

(iii) Dominant energy condition imposes that $\varrho\ge |p_{i}|$, $(i=1,2,3)$
which is illustrated in Fig. \ref{figD}. In fact, it yields the following relation,
\begin{equation}
D\equiv1-\frac{6r^2}{l^2}+(1+\frac{2r}{3}+\frac{r^2}{18})Be^{\frac{r}{3}}\geq0.
\end{equation}
It is clear that this condition is satisfied for $r\ll r_{h}$ if $B\approx l$. But, for the cases of $B\gg l$ the satisfaction of this condition is unrestricted.

So all the energy conditions may be satisfied together if
$B\gg l$. Using these restrictions
of parameter values, all the energy conditions are satisfied on the
black hole horizon. These conditions are similar to the energy
conditions for Chaplygin black hole \cite{Debnath1}. For Van der
Waals black hole \cite{Del, van1}, some of the energy conditions
are violated but for polytropic black hole \cite{Debnath1,Kanzi:2021jrl}, all the
energy conditions are satisfied. In our modified cosmic Chaplygin
black hole, some of the energy conditions are satisfied for some
restrictions of the parameters $B\gg l$.

\begin{figure}[h!]
 \begin{center}$
 \begin{array}{cccc}
\includegraphics[width=55 mm]{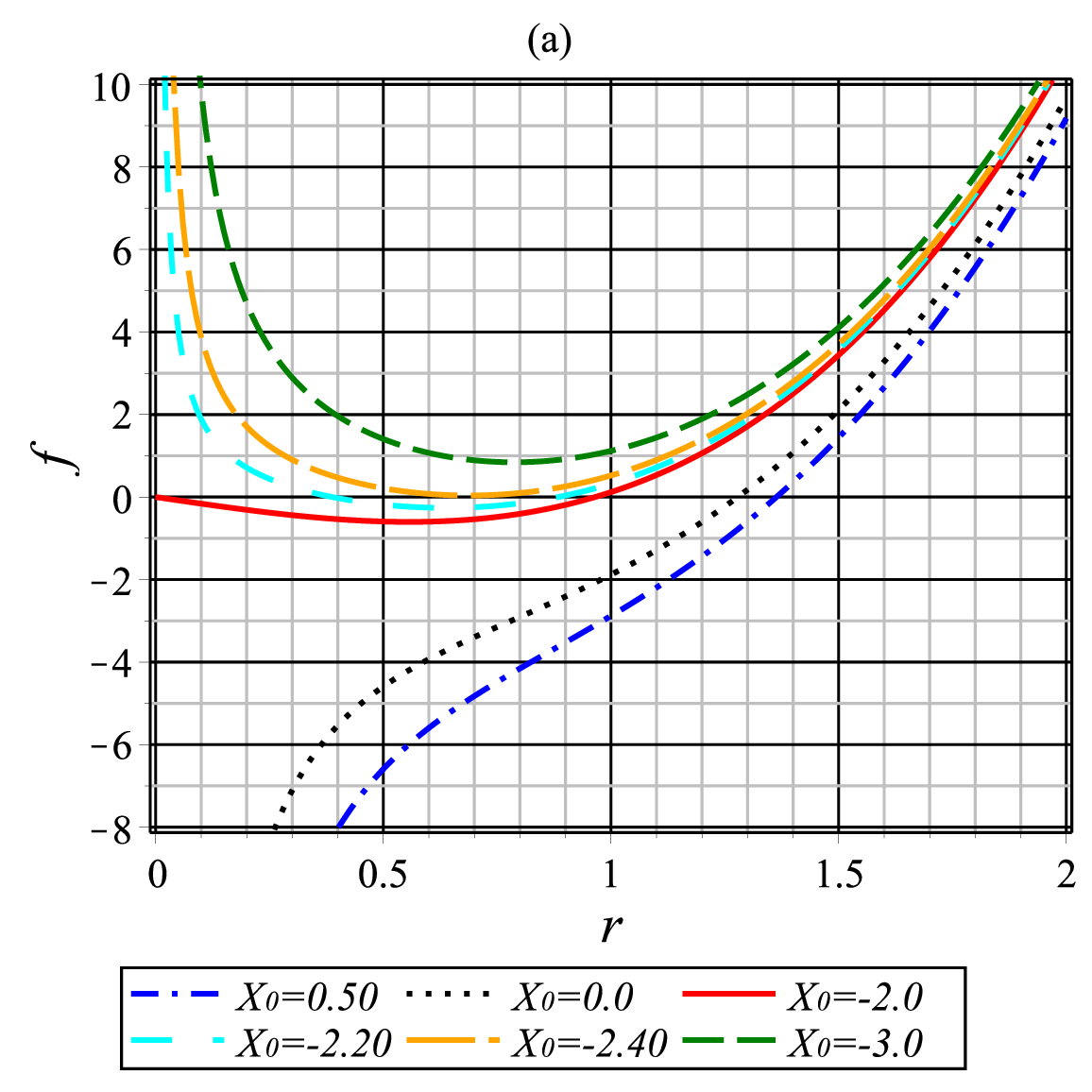}\includegraphics[width=55 mm]{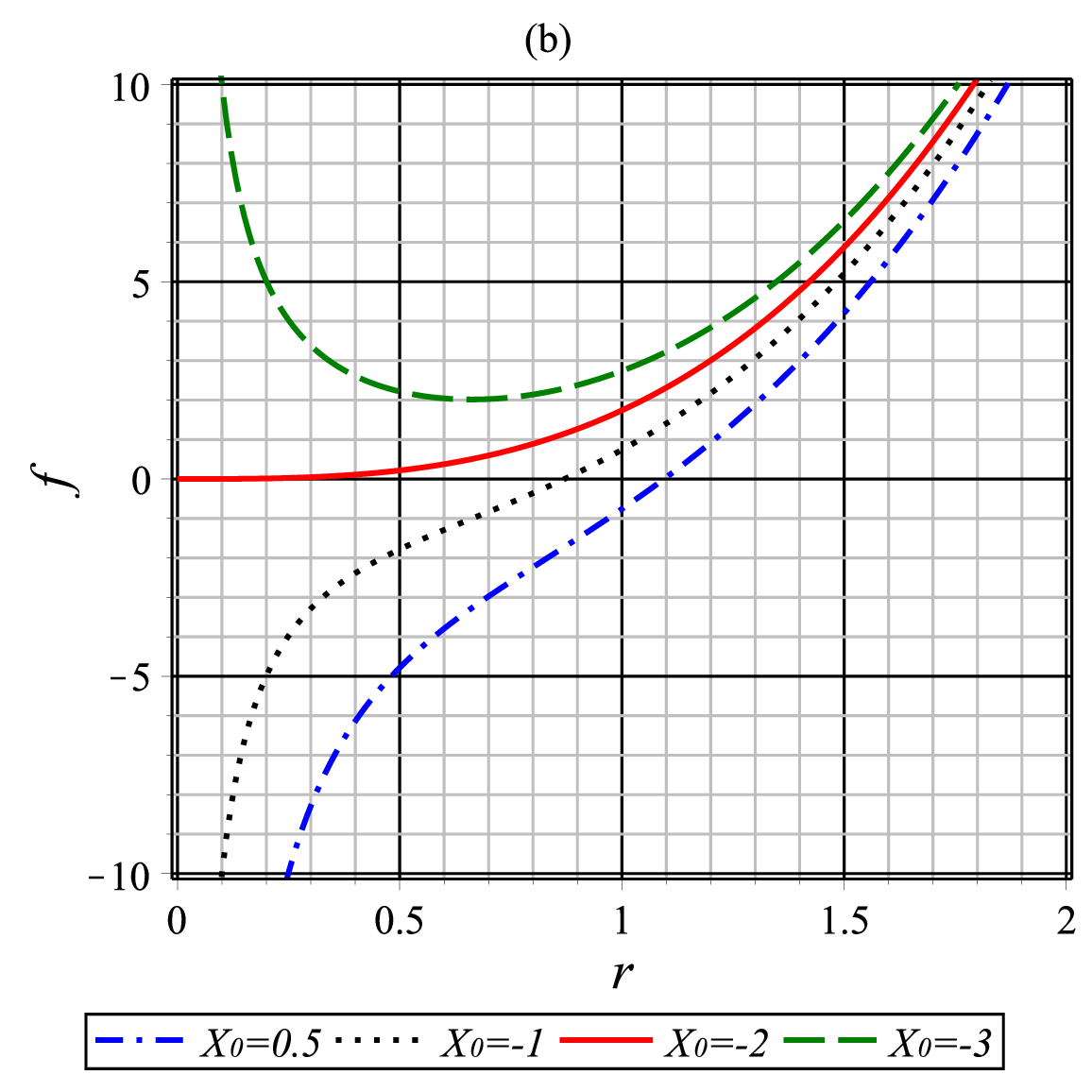}\includegraphics[width=55 mm]{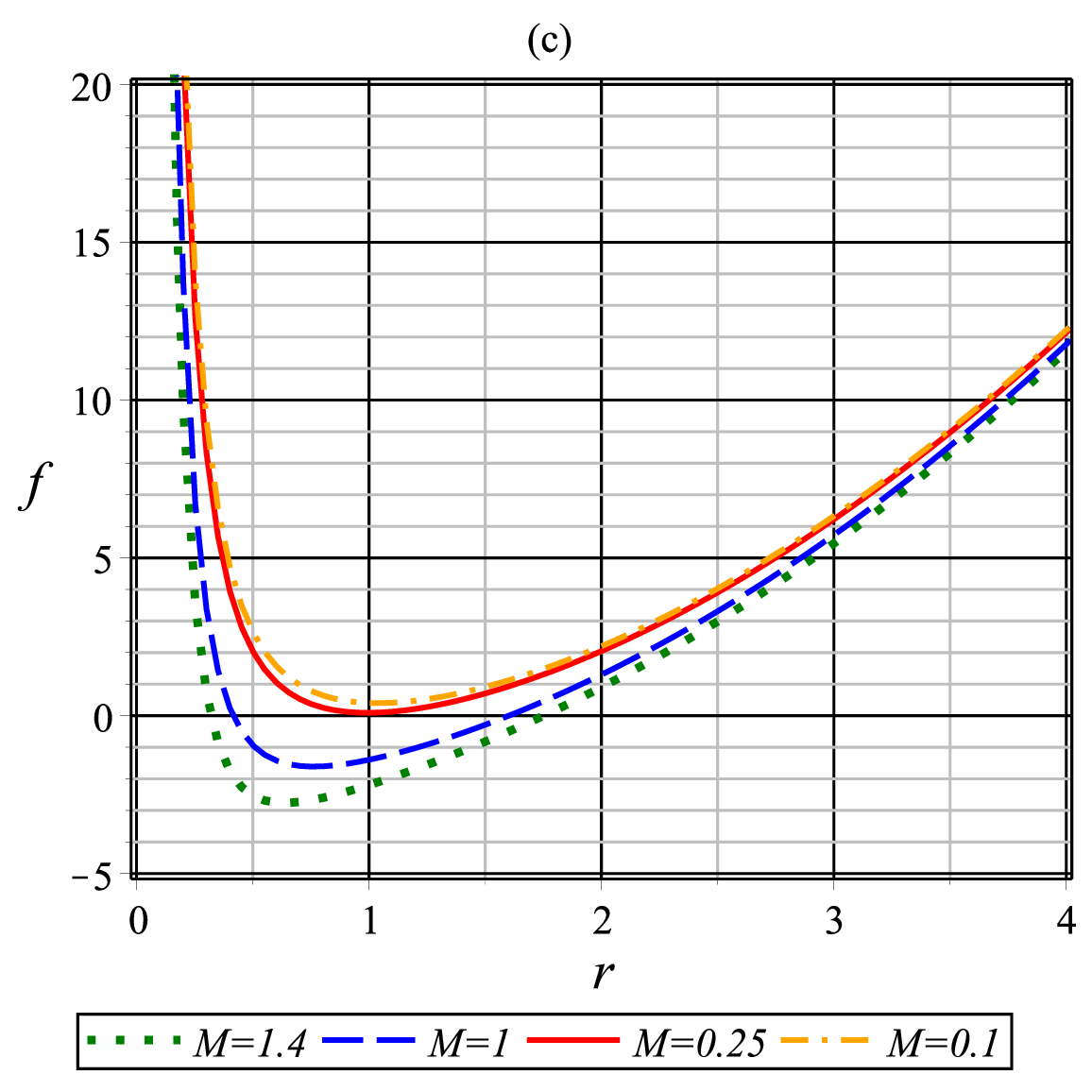}
 \end{array}$
 \end{center}
\caption{Horizon structure of MCCG AdS black hole for $l=1$ (a) $M=1$, $Q=1$  (b) $B=1$, $M=1$  (c) $B=1$, $Q=1$.}
 \label{fig2}
\end{figure}

Now, we can analyze the horizon structure of the MCCG AdS black hole. We find that there is at least one positive real root by choosing suitable values for the parameters of the model. We can see horizon structure for several model parameters by the plots of the Fig. \ref{fig2}. In the Fig. \ref{fig2} (a) we fix $M$ and $Q$ to see effect of the constant $B$. In the cases of $B\approx l$ there are inner and outer horizons. While larger $B$ yields to having three roots or even one root. Extremal case or naked singularity depends on values of $M$ and $Q$. In the Fig. \ref{fig2} (b) we fix $M$ and $B$ to see that by suitable choice of $Q$ one can obtain regular black hole as well as naked singularity. In the case of uncharged black hole ($Q=0$) there is only one root. In the Fig. \ref{fig2} (c) we fix $Q$ and $B$ to see that black hole mass parameter is important to get extremal case (solid red line) or naked singularity (dash dotted orange line of the Fig. \ref{fig2} (c)). Larger black hole mass parameter yields to two real positive roots which is illustrated by dotted green and dashed blue lines of the Fig. \ref{fig2} (c). It should be noted that the values of the black hole parameters are all set to unity in Fig. \ref{fig2}, however the general behavior is not significantly affected by the variation of those parameters.

\section{Thermodynamics} \label{sec4}
In the previous section, section \ref{sec3}, we discussed about the horizon radius $r_{h}$, which is obtained from $f(r_{h},\rho)=0$:
\begin{equation}\label{T1}
\frac{r^{2}}{l^{2}}-\frac{2M}{r}+\frac{Q^{2}}{r^{2}}-C_{XY}\exp{\left(\frac{X_{0}r}{3r_{h}}+\frac{8\pi l^{2}(X_{0}+3)r_{h}^{4}}{27(Q^{2}l^{2}+r_{h}^{4})}p\right)}=0,
\end{equation}
where $p$ is given by equation (\ref{3}).
As we discussed in the previous section, $r_{h}$ depends on $M$, $Q$, and $B$. In the plots of Fig. \ref{fig2}, we show some values of $r_{h}$. By using equation (\ref{3}) in equation (\ref{T1}), we can obtain a relation between energy density and horizon radius,
\begin{equation}\label{33}
A\rho-\rho^{-\alpha}\left(C+(\rho^{1+\alpha}-C)^{-w} \right)=\frac{27(Q^{2}l^{2}+r_{h}^{4})}{8\pi l^{2}(3+r_{h})r_{h}^{4}}
\left[\frac{1}{C_{XY}}\ln{(\frac{r_{h}^{2}}{l^{2}}-\frac{2M}{r_{h}}+\frac{Q^{2}}{r_{h}^{2}})}-\frac{r_{h}}{3}\right].
\end{equation}
Now, from equation (\ref{6}), we obtain $r_{h}=\sqrt{\frac{S}{\pi}}$.
Therefore, using equation (\ref{8}), we write down the volume as,
\begin{equation}
V=\frac{4}{3}\sqrt{\frac{S^{3}}{\pi}}-\frac{C_{XY}}{2}\sqrt{\frac{S}{\pi}}\gamma e^{(\frac{1}{3}\sqrt{\frac{S}{\pi}}+\gamma p)},
\end{equation}
where
\begin{equation}
\gamma=\frac{8\pi l^{2}(3+\sqrt{\frac{S}{\pi}})S^{2}}{27(\pi^{2}Q^{2}l^{2}+S^{2})}.
\end{equation}
Also from equation (\ref{9}), we get the temperature
\begin{equation}\label{T2}
T=\frac{1}{4\pi}\left[\frac{2r_{h}}{l^{2}}+\frac{2M}{r_{h}^{2}}-\frac{2Q^{2}}{r_{h}^{3}}-\frac{C_{XY}}{3}e^{\frac{r_{h}}{3}+\gamma p}\right],
\end{equation}
where
\begin{equation}\label{M2}
M=\frac{4\pi}{3}pr_{h}^3+\frac{Q^{2}}{2r_{h}}-\frac{r_{h}}{2}C_{XY}e^{\frac{r_{h}}{3}+\gamma p}
\end{equation}

\begin{figure}[h!]
 \begin{center}$
 \begin{array}{cccc}
\includegraphics[width=70 mm]{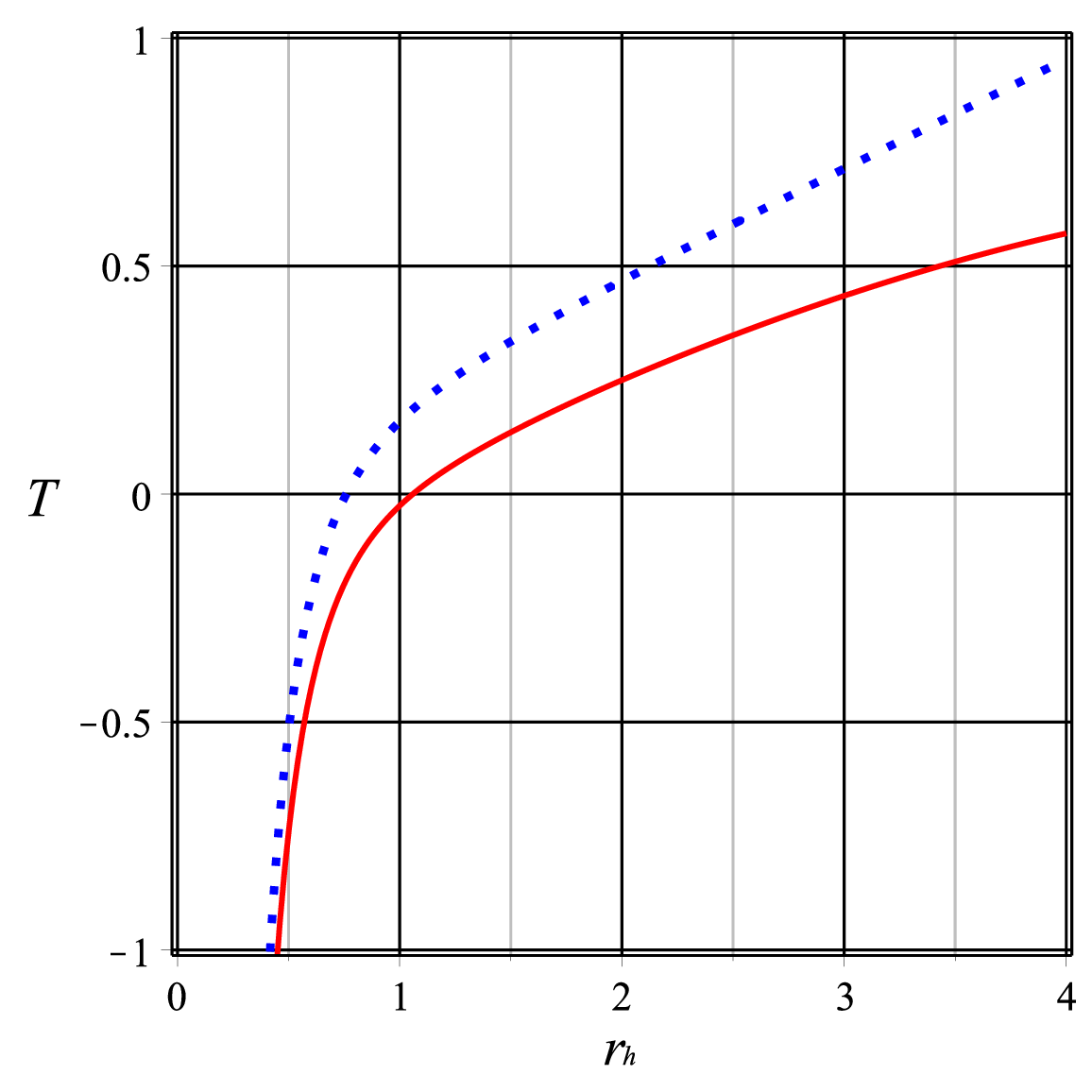}
 \end{array}$
 \end{center}
\caption{Typical behavior of MCCG AdS black hole temperature for $l=1$.}
 \label{fig3}
\end{figure}

\begin{figure}[h!]
 \begin{center}$
 \begin{array}{cccc}
\includegraphics[width=70 mm]{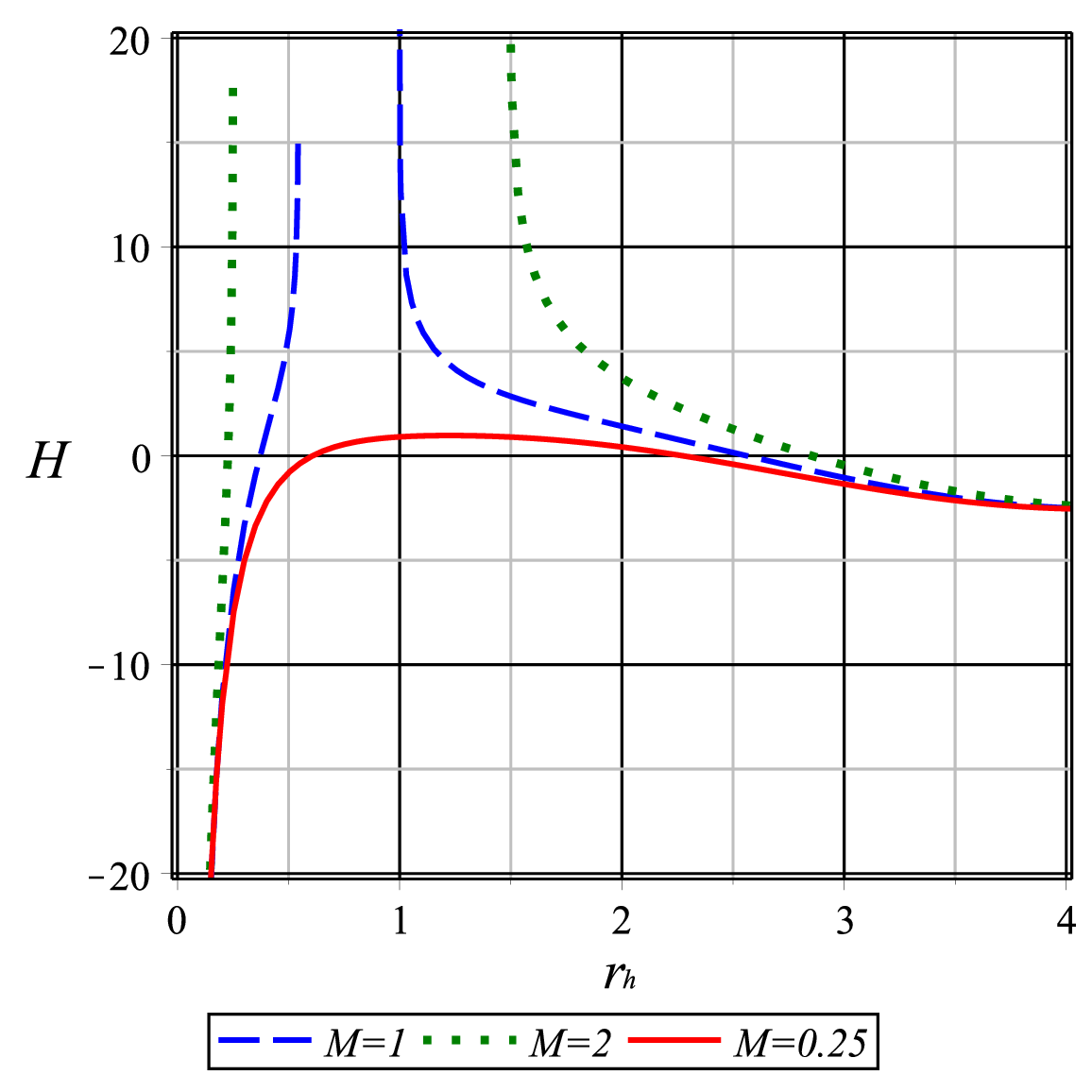}
 \end{array}$
 \end{center}
\caption{Typical behavior of MCCG AdS black hole enthalpy for $l=1$, $Q=1$, and $C_{XY}=1$.}
 \label{figH}
\end{figure}

Assuming $l=1$ and putting the relation (\ref{M2}) in the equation (\ref{T2}) we can obtain,
\begin{equation}\label{T3}
T=\frac{9r_{h}^{4}-3Q^{2}-C_{XY}r_{h}^{2}(3+r_{h})\exp{(\frac{4r_{h}^{5}+3r_{h}^{4}+3r_{h}Q^{2}}{9(Q^{2}+r_{h}^{4})})}}{12\pi r_{h}^{3}},
\end{equation}
where equation (\ref{000}) is used. In the plots of Fig. \ref{fig3}, we draw typical behavior of the temperature in terms of horizon radius. We can see that there is a minimum radius where below that the temperature is negative, hence the black hole is in unstable phase.\\
Now, we want to discuss the thermodynamic behavior of the MCCG AdS black hole in the
presence of variable pressure $p$ (variable cosmological
constant $\Lambda$). The enthalpy
function is defined by $H = U + pV+\Phi Q$ \cite{HL4,Kastor}. Hence, from the first law of
thermodynamics, we can write
\begin{equation}
dH=TdS-Vdp+Qd\Phi.
\end{equation}
Now integrating, the enthalpy function $H$ can be expressed in the following
form,
\begin{eqnarray}
H&=&\left(2r_{h}\ln{(\frac{r_{h}^{4}-2Mr_{h}+Q^{2}}{r_{h}^{2}})}-C_{XY}r_{h}^{2}\right)(\frac{r_{h}^{4}-2Mr_{h}+Q^{2}}{r_{h}^{2}})^{\frac{1}{C_{XY}}}\nonumber\\
&-&\frac{18(r_{h}^{4}+Q^{2})\ln{(\frac{r_{h}^{4}-2Mr_{h}+Q^{2}}{r_{h}^{2}})}-
2C_{XY}(4r_{h}^{5}+3r_{h}^{4}+Mr_{h}^{2}+(4Q^{2}+3M)r_{h}+3Q^{2})}{C_{XY}(3+r_{h})r_{h}}.
\end{eqnarray}
Our graphical analysis, as illustrated by the Fig. \ref{figH}, show a disconnected region when having larger black hole mass parameter for $r_{h}=1$ where the temperature becomes negative. Thus, However, in the extremal black hole (see the Fig. \ref{fig2} (c)) or naked singularity  there is no such discontinuity. The maximum enthalpy is achieved at the event horizon. Remember form the plots of the Fig. \ref{fig2} that for selected values of parameters we have $r_{h}\approx1.5$.

The Gibbs' free energy  ($G=H-TS$) can be expressed in the form of
\cite{Graca}:
\begin{eqnarray}
G&=&\left(\frac{r_{h}}{2}\ln{(\frac{r_{h}^{4}-2Mr_{h}+Q^{2}}{r_{h}^{2}})}-\frac{C_{XY}r_{h}^{2}}{6}\right)(\frac{r_{h}^{4}-2Mr_{h}+Q^{2}}{r_{h}^{2}})^{\frac{1}{C_{XY}}}\nonumber\\
&-&\frac{9(r_{h}^{4}+Q^{2})\ln{(\frac{r_{h}^{4}-2Mr_{h}+Q^{2}}{r_{h}^{2}})}-
5C_{XY}(\frac{3}{5}r_{h}^{5}+Q^{2}(r_{h}+\frac{6}{5}))}{2C_{XY}(3+r_{h})r_{h}}.
\end{eqnarray}
General behavior of the Gibbs energy is similar to the enthalpy which sounds some instability. In order to confirm such instability, we should analyze the Helmholtz free energy ($F=G-pV$) which is obtained by \cite{Graca},
\begin{eqnarray}
F&=&\left(r_{h}\ln{(\frac{r_{h}^{4}-2Mr_{h}+Q^{2}}{r_{h}^{2}})}-\frac{C_{XY}r_{h}^{2}}{3}\right)(\frac{r_{h}^{4}-2Mr_{h}+Q^{2}}{r_{h}^{2}})^{\frac{1}{C_{XY}}}\nonumber\\
&-&\frac{9(r_{h}^{4}+Q^{2})\ln{(\frac{r_{h}^{4}-2Mr_{h}+Q^{2}}{r_{h}^{2}})}-
4C_{XY}(\frac{3}{4}r_{h}^{5}+Q^{2}(r_{h}+\frac{3}{4}))}{C_{XY}(3+r_{h})r_{h}}.
\end{eqnarray}

\begin{figure}[h!]
 \begin{center}$
 \begin{array}{cccc}
\includegraphics[width=70 mm]{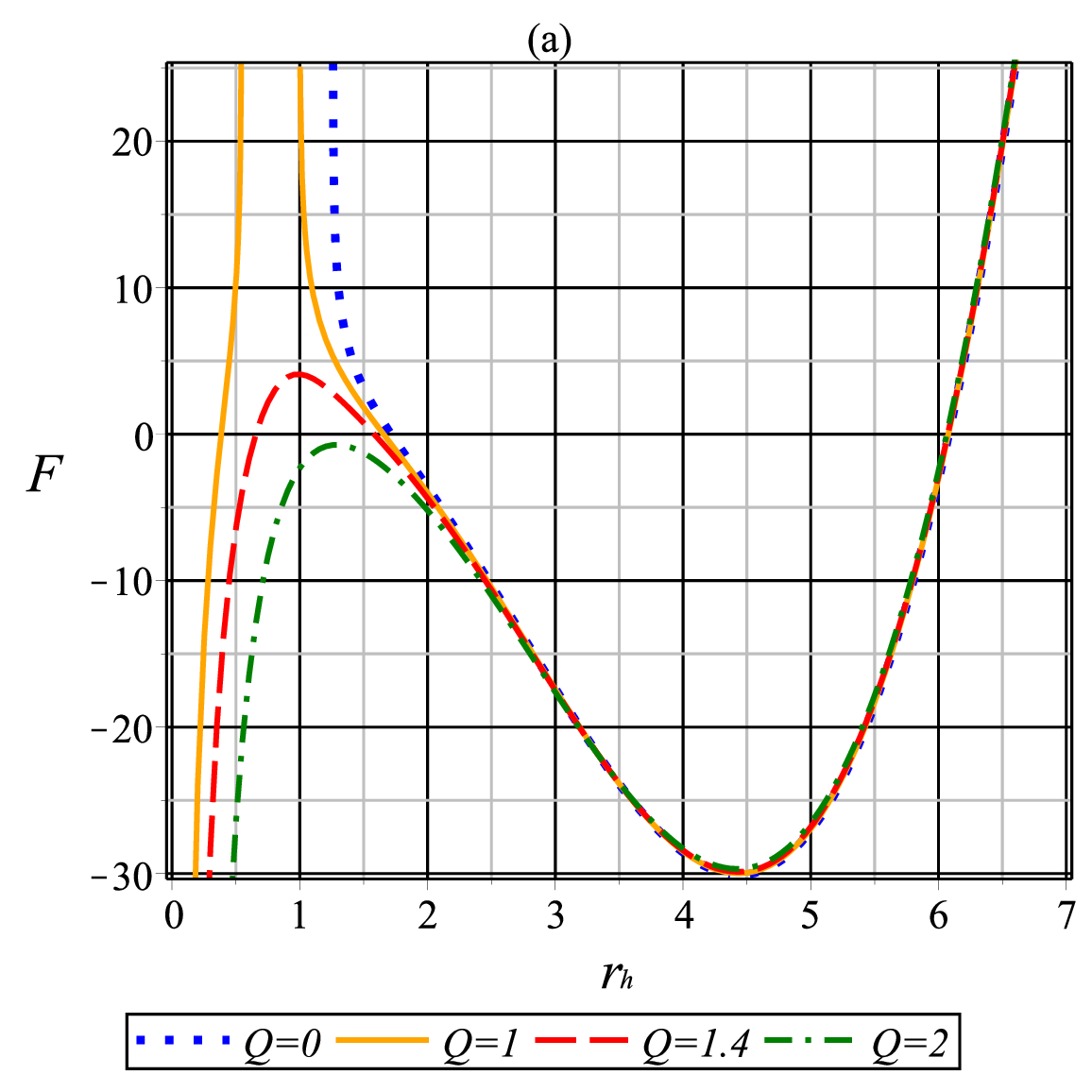}\includegraphics[width=70 mm]{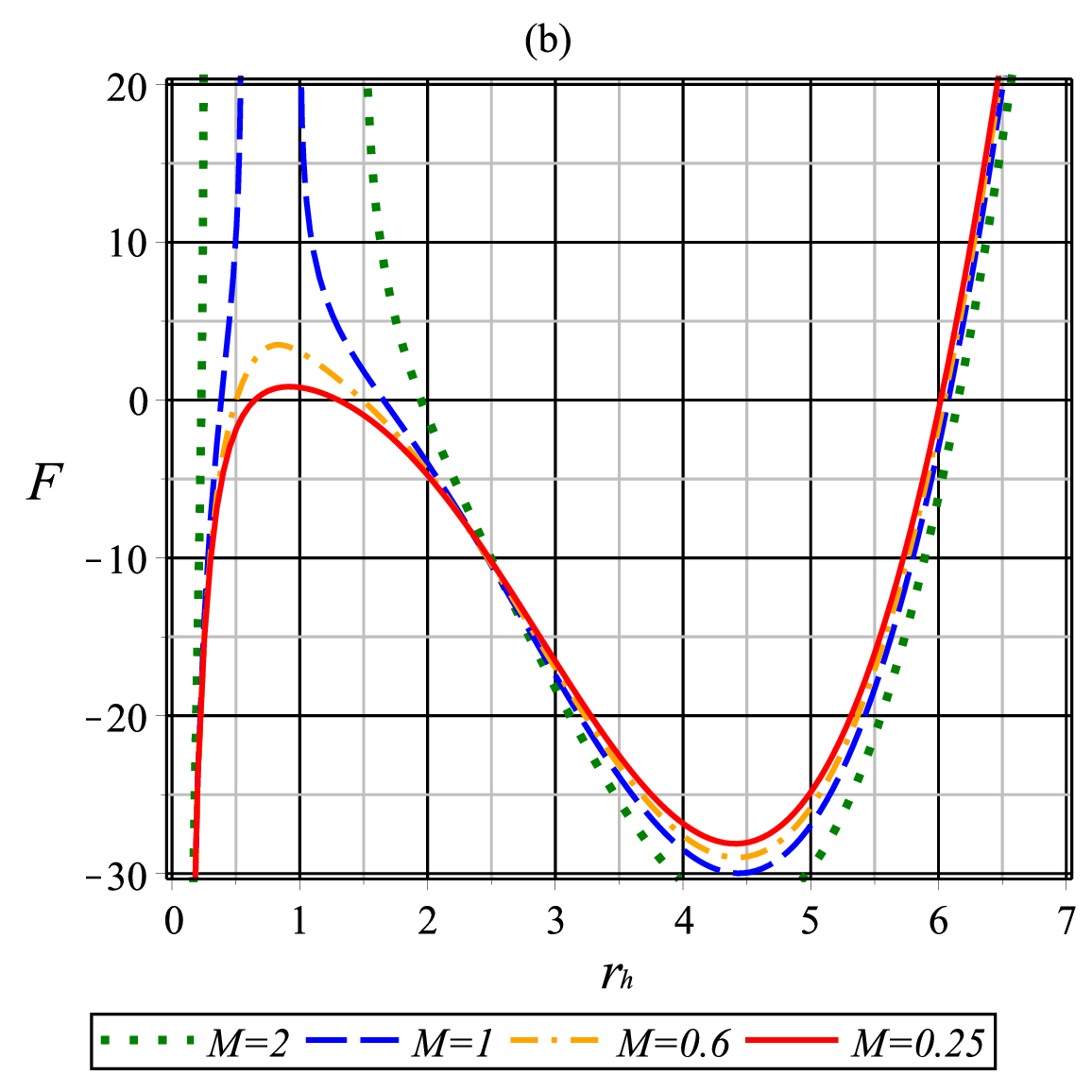}
 \end{array}$
 \end{center}
\caption{Typical behavior of Helmholtz free energy for $l=1$ and $C_{XY}=1$ (a) $M=1$; (b) $Q=1$.}
 \label{fig4}
\end{figure}

\begin{figure}[h!]
 \begin{center}$
 \begin{array}{cccc}
\includegraphics[width=65 mm]{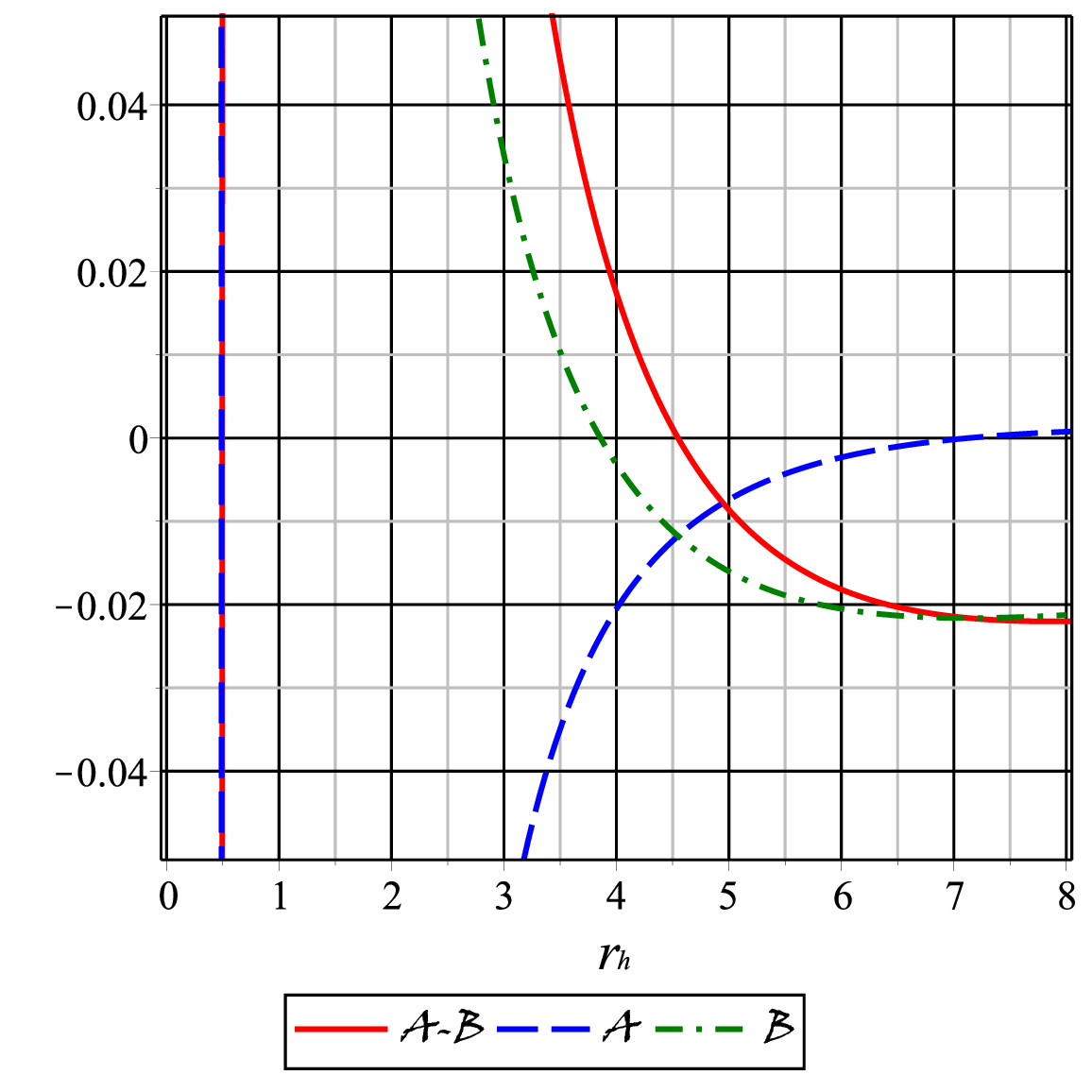}
 \end{array}$
 \end{center}
\caption{Critical points analysis for $l=1$, $C_{XY}=1$, $M=1$ and $Q=1$.}
 \label{Crit}
\end{figure}

In the plots of the Fig. \ref{fig4} we draw Helmholtz free energy to see typical behavior.
General behavior of the Helmholtz free energy illustrated by plots of the Fig. \ref{fig4}. We can see that there is not a global minimum which may be sign of some instability of the model. For the large $Q$ or small $M$ the Helmholtz free energy is continues with local extremum points (one minimum and one maximum) which may be aware us a phase transition. We will discuss about that by analyzing specific heat. Before that we briefly discuss about the critical points.\\
For the charged AdS black holes, the critical behavior has been found
by the Ref. \cite{Kubi}. The critical point is an inflection point
which can be found by the following conditions,
\begin{eqnarray}
\mathcal{A}&=&\left(\frac{\partial p}{\partial r_{h}}\right)_{cr}=0,\nonumber\\
\mathcal{B}&=&\left(\frac{\partial^{2} p}{\partial r_{h}^{2}}\right)_{cr}=0.
\end{eqnarray}
Combining the equations (\ref{3}) and (\ref{33}) we find that there is no such points for charged black hole as well as uncharged black holes, it is illustrated by Fig. \ref{Crit}. Dash dotted green line of the Fig. \ref{Crit} shows $\mathcal{A}$ which is equal to $\mathcal{B}$ (dashed blue line of the Fig. \ref{Crit}) at $r_{cr}$. Solid red line of the Fig. \ref{Crit} confirms that at $r_{h}=r_{cr}$ $\mathcal{A}=\mathcal{B}$ ($\mathcal{A}-\mathcal{B}=0$), however vales of $\mathcal{A}$ and $\mathcal{B}$ are not zero at overlapped point hence there is no any critical points.

\subsection{Stability}
We can discuss about the stability of the MCCG AdS black hole by analyzing specific heat capacity which is given
by,
\begin{equation}\label{c1}
{\cal C}=\frac{\left(2r_{h}+\frac{2M}{r_{h}^{2}}-\frac{2Q^{2}}{r_{h}^{3}}-\frac{C_{XY}}{3}(r_{h}^{2}-\frac{2M}{r_h}+\frac{Q^{2}}{r_{h}^{2}})^{\frac{1}{C_{XY}}}\right)\pi r_{h}}{1-\frac{2M}{r_{h}^{3}}+\frac{3Q^{2}}{r_{h}^{4}}-\frac{(r_{h}+\frac{M}{r_{h}^{2}}-\frac{Q^{2}}{r_{h}^{3}})(r_{h}^{2}-\frac{2M}{r_h}+\frac{Q^{2}}{r_{h}^{2}})^{\frac{1}{C_{XY}}}}{3(r_{h}^{2}-\frac{2M}{r_{h}}
+\frac{Q^{2}}{r_{h}^{2}})}}.
\end{equation}
If ${\cal C}>0$, the black hole is stable, and if ${\cal C}<0$,
the black hole is in an unstable phase. In the Fig. \ref{5} we draw specific heat and see that there is first and second order phase transition. In the Fig. \ref{5} (a) we fix $C_{XY}$ and $Q$ to see effect of the black hole mass parameter. For the smaller mass parameter we can see the first order phase transition which show stable/unstable phase transition. For the larger mass parameter (as illustrated by solid red line of the Fig. \ref{5}) we can see also second order phase transition. The system with the small horizon radius is in the stable phase.\\
In the Fig. \ref{5} (b) we fix $M$ and $C_{XY}$ to see effect of the black hole charge. In the case of uncharged black hole, system is mostly in unstable phase. However, we show that charged black hole with small horizon radius is in stable phase with possibility of a phase transition.\\
Finally in the Fig. \ref{5} (c) we fix $M$ and $Q$ to see effect of the $C_{XY}$. We can see possible phase transition of this case as well as the previous cases.

\begin{figure}[h!]
 \begin{center}$
 \begin{array}{cccc}
\includegraphics[width=55 mm]{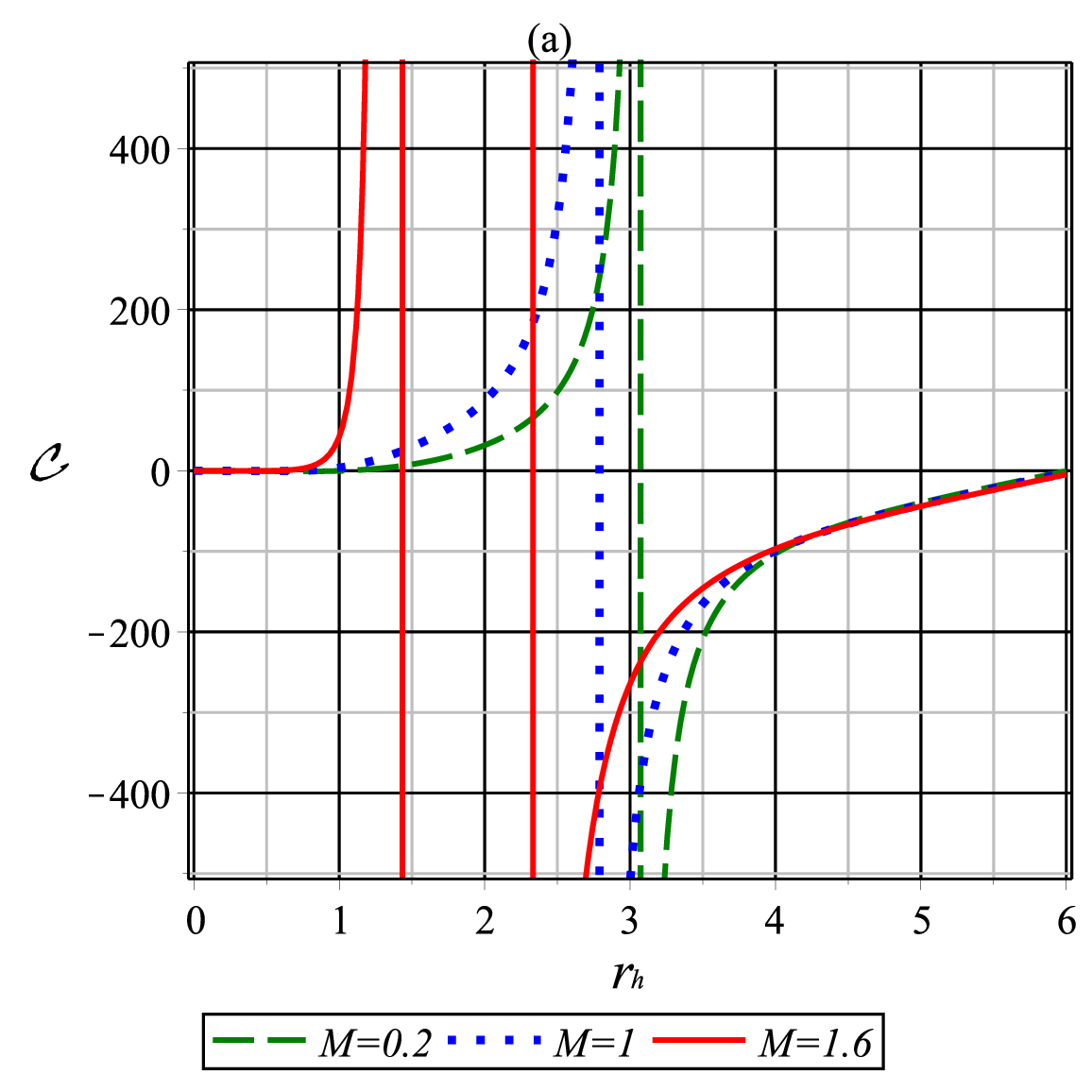}\includegraphics[width=55 mm]{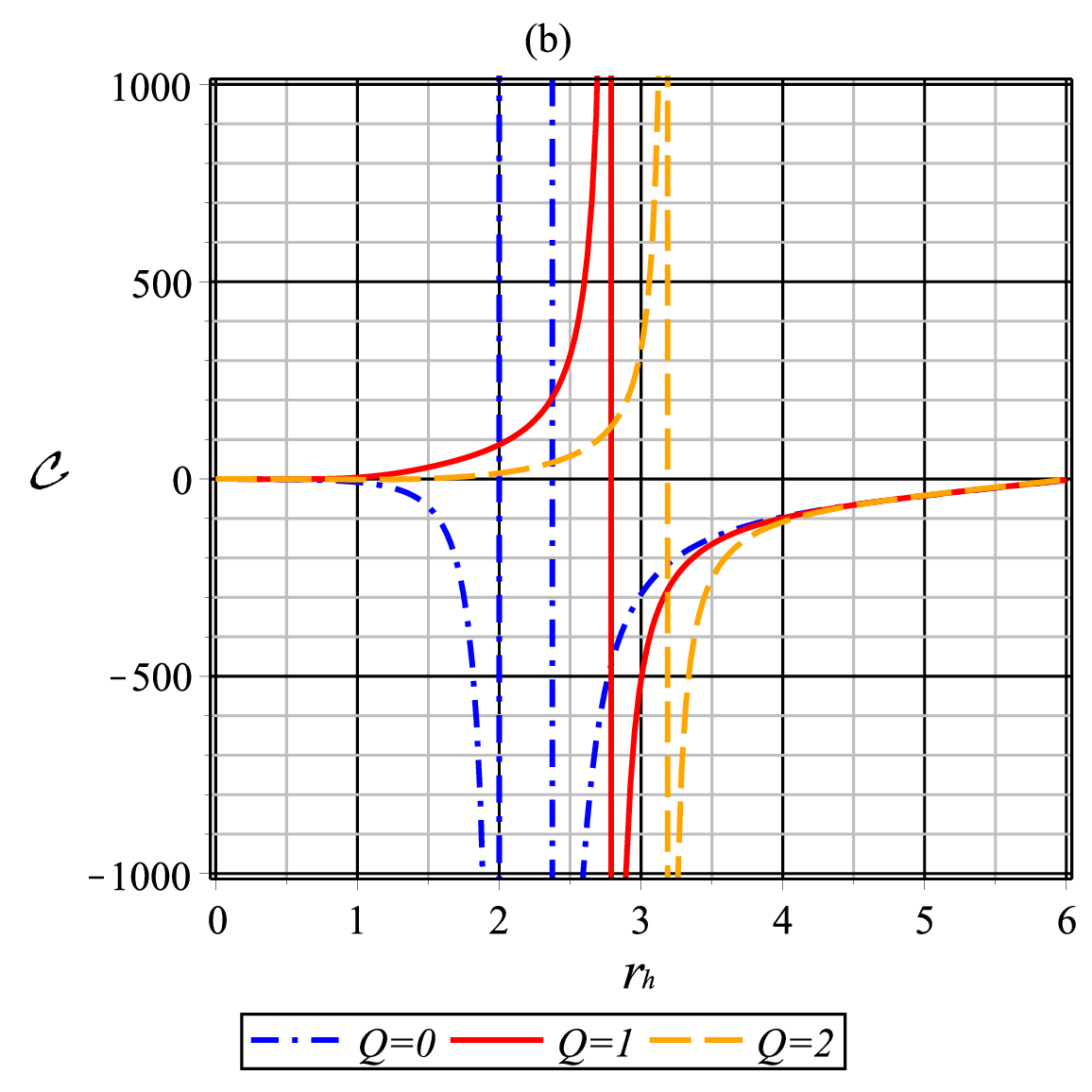}\includegraphics[width=55 mm]{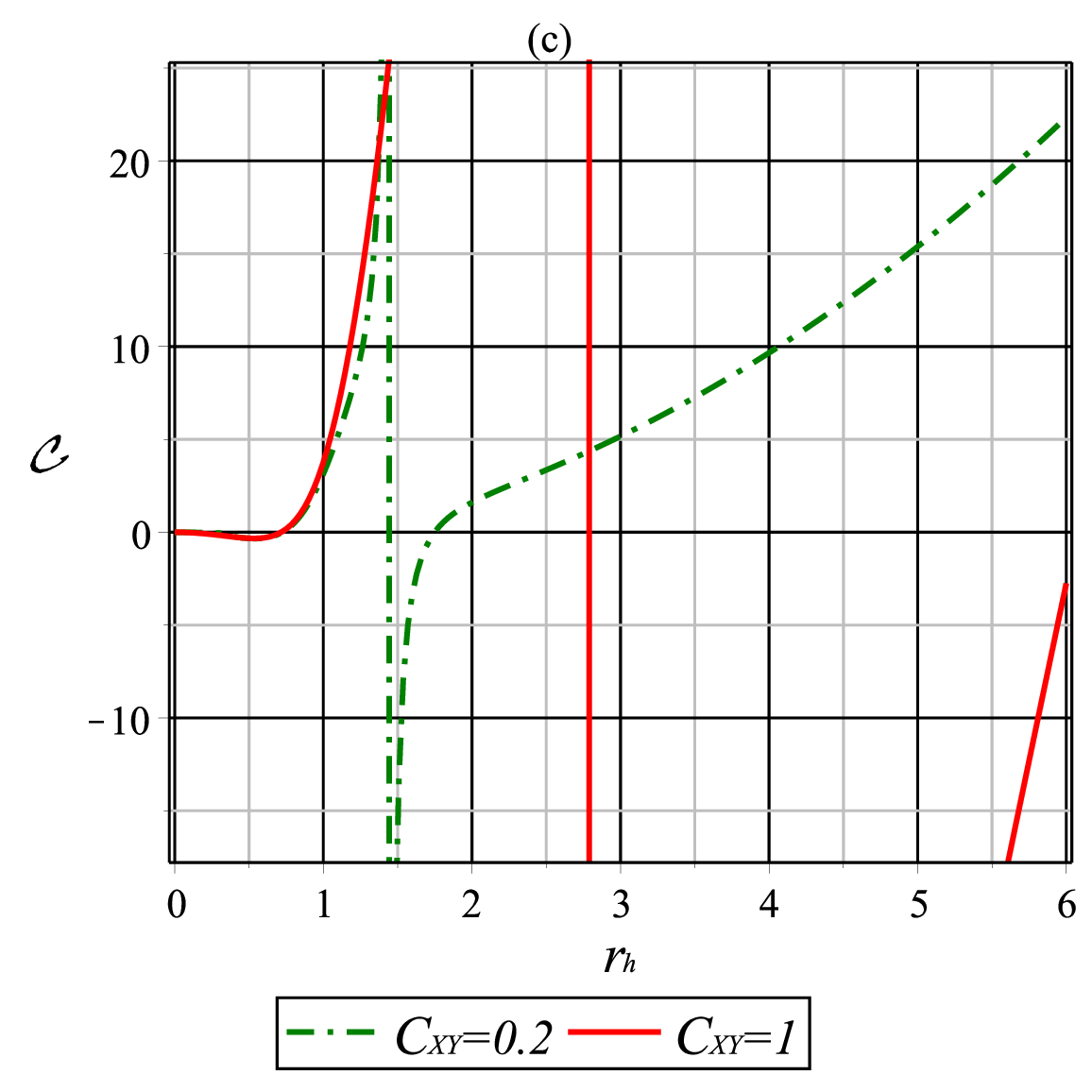}
 \end{array}$
 \end{center}
\caption{Typical behavior of specific heat for $l=1$. (a) $C_{XY}=1$, $Q=1$ (b) $C_{XY}=1$, $M=1$ (c) $Q=1$, $M=1$.}
 \label{fig5}
\end{figure}

\subsection{Joule-Thompson expansion}
\begin{figure}[h!]
 \begin{center}$
 \begin{array}{cccc}
\includegraphics[width=65 mm]{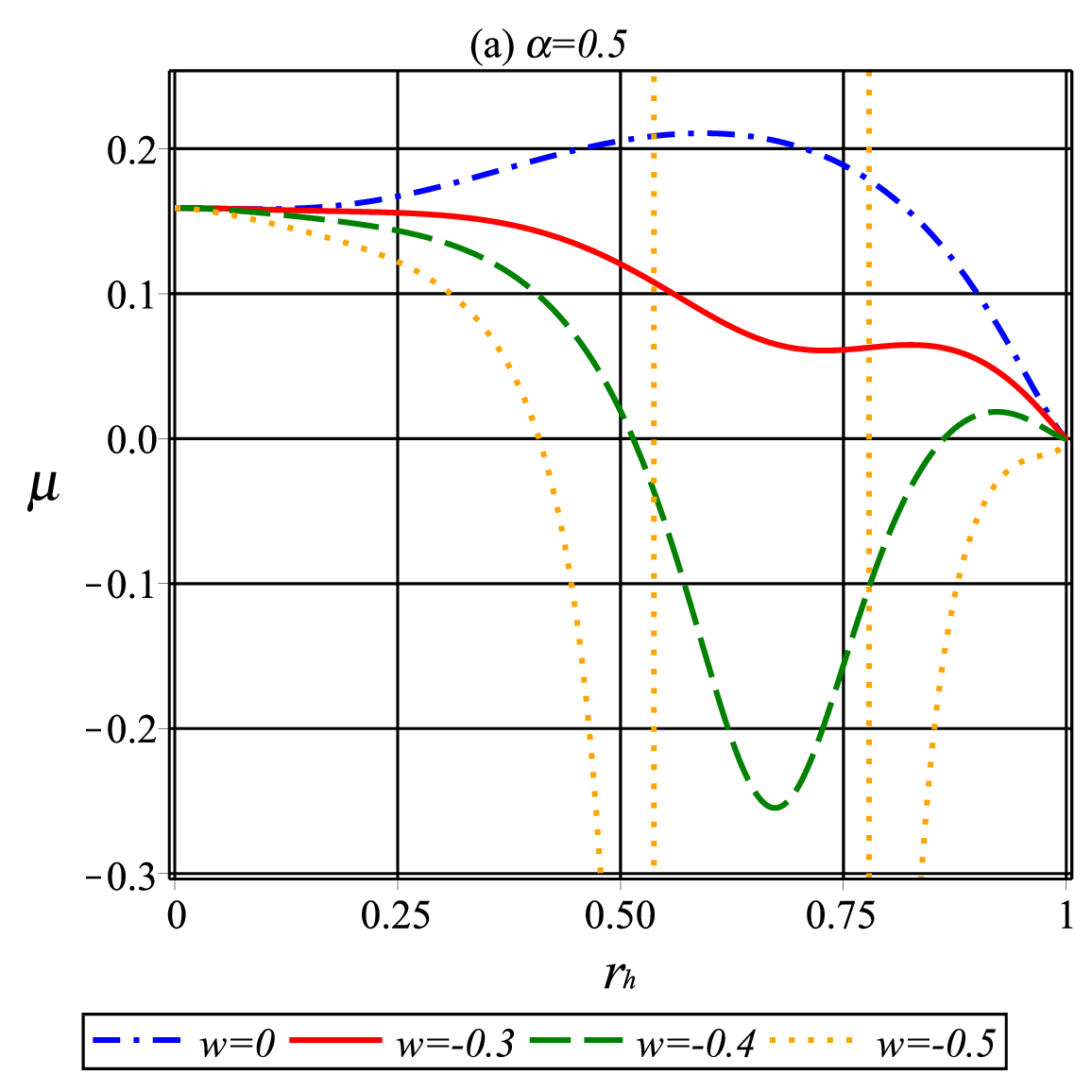}\includegraphics[width=65 mm]{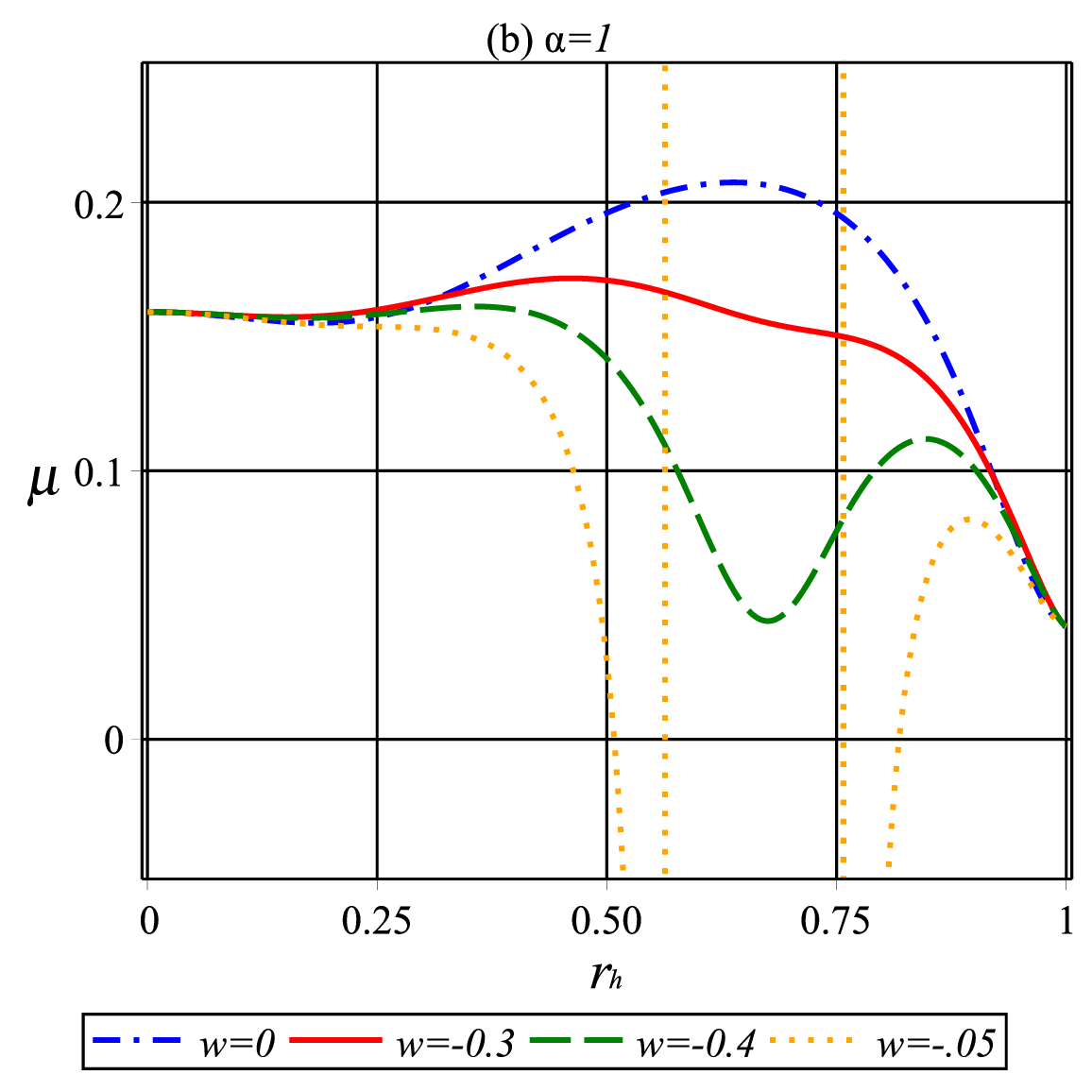}
 \end{array}$
 \end{center}
\caption{Joule-Thomson coefficient in terms of horizon radius for $M=1$ and $l=1$. (a) $Q=1$; (b) $C_{XY}=1$.}
 \label{fig6}
\end{figure}

Joule-Thomson expansion is determined by changing the temperature
with respect to the pressure, while the enthalpy remains constant. Since in the AdS space, the black
hole mass is interpreted as enthalpy \cite{Kastor}, so during the
Joule-Thomson expansion process, the mass of the black hole
remains constant. The Joule-Thomson coefficient is given by
\cite{Riz,Ros}
\begin{equation}
\mu=\left(\frac{\partial T}{\partial p}\right)_{M}=\frac{1}{{\cal
C}}\left[T\left(\frac{\partial V}{\partial T}\right)_{p}-V
\right].
\end{equation}
It is possible to determine whether cooling or heating will occur
by evaluating the sign of $\mu$. In Joule-Thomson expansion, the
pressure decreases i.e., the change of pressure is negative but the
change of temperature may be positive or negative. If the change
of temperature is positive (negative) then $\mu$ is negative
(positive) and therefore heating (cooling) occurs.\\
Graphically, we obtain behavior of Joule-Thomson coefficient with the event horizon radius which is illustrated by the Fig. \ref{fig6}

Both heating and cooling are found to be dependent on the horizon radius in the MCCG AdS black hole. In the case of $M=l=1$ we draw Joule-Thomson coefficient in terms of horizon radius in the Fig. \ref{fig6}. However, in the case of uncharged black hole the sign of $\mu$ is completely negative for $r_{h}>r_{c}$ (see solid blue line of Fig. \ref{fig6} (b)), hence the change of temperature is positive, which means the black hole is unstable as expected.\\
In the case of fixed black hole charge (Fig. \ref{fig6} (a)), and large $C_{XY}$ the value of $\mu$ is positive for small values of $r_{h}$. Larger values of horizon radius yields to negative $\mu$. It means that for the large values of $r_{h}$ there are some unstable regions. However, the situation is significantly more difficult for lower values of $C_{XY}$, as there are multiple cooling and heating zones depending on the value of the horizon radius.\\
We can see similar behavior for the case of $C_{XY}=1$ and varying $Q$ (Fig. \ref{fig6} (b)). Only regular behavior is for the case of uncharged black hole, the stable unstable phase transition happen at special radius we called $r_c$. In the case of charged black hole it seems there are the first and second phase transition as discussed already.

\section{Heat Engine} \label{sec5}
In this section, we study heat engine based on MCCG AdS black hole (uncharged case). The Ref. \cite{John} was the first to introduce a holographic heat engine for a given black hole. Holographic heat engines were then developed by the following Refs. \cite{Joh,Joh1,Joh2}. Dilatonic Born-Infeld and $f(R)$ black holes were also proposed as the candidates of the holographic heat engine \cite{Bha,Zhan}. Static and dynamical black holes as the heat engines were considered in Ref. \cite{Sade}, which, in particular, studied the effects of dynamical parameters (rotation and charges) on the thermodynamic efficiency of the holographic heat engine. Ref. \cite{Chakra} compared such efficiency of different working substances. BTZ black holes are important kinds of black holes \cite{BTZ1, BTZ2, BTZ3}. For example in the Ref. \cite{BTZ3} heat engine of BTZ black hole considered. Also a massive charged
BTZ black hole as a heat engine constructed by the Ref. \cite{Hen}. Already, heat engine in the 3-dimensional charged BTZ black holes defined \cite{Mo}. In the interesting work \cite{Liu} the effect of dark energy on the efficiency of the Reissner-Nordstr\"{o}m-AdS black holes as heat engine investigated. C. V. Johnson already studied the holographic heat engine and found that 4-dimensional Taub-Bolt-AdS spacetime is similar to the Schwarzschild-AdS black hole \cite{Jo}. There are also several other works where the
heat engine mechanism considered for different kinds of black holes
\cite{Hu1, Mo1, Hendi, Wei1, Avik, Fang, Zhang, Rosso, Moo, Pana, J, Hhu, Santo, Fern, Z}. Recently, Ref. \cite{Set1} have discussed polytropic black hole as a heat engine, and Debnath \cite{Debnath1} has studied the classical heat engine for
Chaplygin black hole. Motivated by these works, here we'll study
the heat engine for our constructed MCCG AdS black hole.

\subsection{Carnot Cycle}
\begin{figure}[h!]
 \begin{center}$
 \begin{array}{cccc}
\includegraphics[width=70 mm]{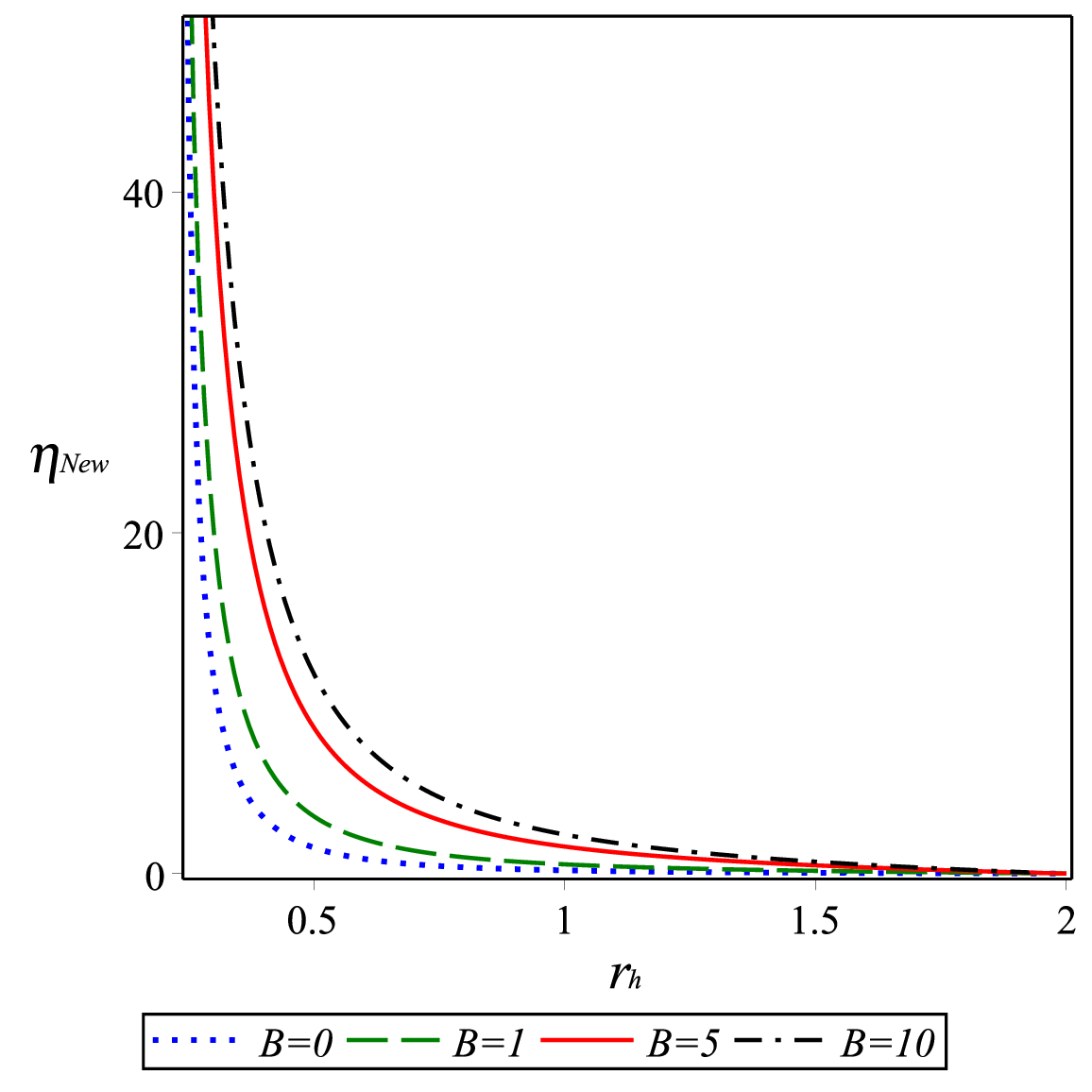}
 \end{array}$
 \end{center}
\caption{Typical behavior of thermal efficiency in terms of horizon radius for $M=2$, $Q=0.2$, $A=1$, $l=1$ and $\alpha=0.5$.}
 \label{fig-eta}
\end{figure}

In 1924, S. Carnot proposed the theoretical ideal thermodynamic
cycle known as Carnot cycle, which provides that any classical
thermodynamic engine can be achieved during the conversion of heat
into work and conversely, the efficiency of a refrigeration system
is creating a temperature difference by the application of work to
the system. A heat engine can act by transferring energy from a
warm region to a cool region and converting some of the energy to
the mechanical work. The Carnot cycle is the simple cycle that
contains two temperatures: a heat source and a heat sink. Now
assume, $T_{H}$ and $T_{C}$ are respectively the temperatures of
the hot and cold reservoirs and they include two isothermal
processes with two adiabatic processes. The Carnot heat engine forms a
closed path in $p$-$V$ diagram \cite{John}. So for the upper isotherm process (denoted by 1 and 2), the heat flow is given by \cite{John},
\begin{equation}
Q_{H}=T_{H}\triangle S_{1\rightarrow 2}=T_{H}(S_{2}-S_{1}),
\end{equation}
and consequently, the exhausted heat produced from the lower
isothermal process (denoted by 3 and 4) is given by \cite{John},
\begin{equation}
Q_{C}=T_{C}\triangle S_{3\rightarrow 4}=T_{C}(S_{3}-S_{4}).
\end{equation}
Here $S_{i}$'s are related to $V_{i}$'s satisfying the following
expression,
\begin{equation}\label{37}
V_{i}=\frac{4}{3}\sqrt{\frac{S_{i}^{3}}{\pi}}-\frac{C_{XY}}{2}\sqrt{\frac{S_{i}}{\pi}}\gamma_{i} e^{(\frac{1}{3}\sqrt{\frac{S_{i}}{\pi}}+\gamma_{i} p_{i})},
\end{equation}
where $i=1,2,3,4$ and
\begin{equation}\label{37-1}
\gamma_{i}=\frac{8\pi l^{2}(3+\sqrt{\frac{S_{i}}{\pi}})S_{i}^{2}}{27(\pi^{2}Q^{2}l^{2}+S_{i}^{2})}.
\end{equation}
Also
\begin{equation}\label{38}
p_{i}=A\rho_{i}-\rho_{i}^{-\alpha}\left(C+(\rho_{i}^{1+\alpha}-C)^{-w}
\right)~,~i=1,2,3,4,
\end{equation}
which is calculated as,
\begin{equation}\label{333}
p_{i}=\frac{27(Q^{2}l^{2}\pi^{2}+S_{i}^{2})}{8\pi l^{2}(3+\sqrt{\frac{S_{i}}{\pi}})S_{i}^{2}}
\left[\frac{1}{C_{XY}}\ln{(\frac{S_{i}}{\pi l^{2}}-\frac{2M\sqrt{\pi}}{\sqrt{S_{i}}}+\frac{\pi Q^{2}}{S_{i}})}-\frac{1}{3}\sqrt{\frac{S_{i}}{\pi}}\right].
\end{equation}
The work done by the heat engine is defined by
\begin{equation}
W=Q_{H}-Q_{C}
\end{equation}
The efficiency of a heat engine (Carnot engine) is defines as the
ratio of work output from the machine and the amount of heat
energy of input at the higher temperature, which is defined by
\begin{equation}
\eta_{_{Car}}=\frac{W}{Q_{H}}=1-\frac{Q_{C}}{Q_{H}}
\end{equation}
For Carnot cycle, $V_{1}=V_{4}$ and $V_{2}=V_{3}$, so the maximum
efficiency for Carnot cycle is,
\begin{equation}
(\eta_{_{Car}})_{max}=1-\frac{T_{C}}{T_{H}}
\end{equation}
which is the maximum one of all the possible cycles between the
given higher temperature $T_{H}$ and lower one $T_{C}$. It should
be mentioned that the Stirling cycle consists of two isothermal
processes and two isochores processes. So the maximally efficient
Carnot engine is also a Stirling engine.\\
There is also a new engine which consists of
two isobars and two isochores/adiabats \cite{John}. In this case, the work done along
the isobars is described by,
\begin{eqnarray}
W=\triangle p_{4\rightarrow 1}~\triangle V_{1\rightarrow
2}=(p_{1}-p_{4})(V_{2}-V_{1}),
\end{eqnarray}
where $V_{1}$, $V_{2}$ are given in (\ref{37}) and $p_{1}$,
$p_{4}$ are given in (\ref{38}). The net inflow of heat in upper
isobar is given by,
\begin{equation}
Q_{H}=\int_{T_{1}}^{T_{2}} {\cal C}(p_{1},T)dT,
\end{equation}
which can be simplified to
\begin{eqnarray}
Q_{H}&\approx&Q^{2}\sqrt{\pi}\left(\frac{1}{\sqrt{S_{2}}}-\frac{1}{\sqrt{S_{1}}}\right)+M\ln{\sqrt{\frac{S_{2}}{S_{1}}}}+\frac{1}{3\sqrt{\pi^{3}}}
\left(\sqrt{S_{2}^{3}}-\sqrt{S_{1}^{3}}\right)\nonumber\\
&-&\frac{C_{XY}^{5}}{12(C_{XY}+1)\pi^{1+\frac{1}{C_{XY}}}}\left(S_{2}^{1+\frac{1}{C_{XY}}}-S_{1}^{1+\frac{1}{C_{XY}}}\right)
-\frac{C_{XY}M\pi^{\frac{1}{2}-\frac{1}{C_{XY}}}}{3(C_{XY}-2)}\left(S_{2}^{\frac{1}{C_{XY}}-\frac{1}{2}}-S_{1}^{\frac{1}{C_{XY}}-\frac{1}{2}}\right)\nonumber\\
&+&\frac{C_{XY}Q^{2}\pi^{1-\frac{1}{C_{XY}}}}{12(C_{XY}-1)}\left(S_{2}^{\frac{1}{C_{XY}}-1}-S_{1}^{\frac{1}{C_{XY}}-1}\right).
\end{eqnarray}
The exhaust of heat from lower isobar is given by,
\begin{equation}
Q_{C}=\int_{T_{3}}^{T_{4}} {\cal C}(p_{4},T)dT,
\end{equation}
which can be simplified to,
\begin{eqnarray}
Q_{C}&\approx&Q^{2}\sqrt{\pi}\left(\frac{1}{\sqrt{S_{4}}}-\frac{1}{\sqrt{S_{3}}}\right)+M\ln{\sqrt{\frac{S_{4}}{S_{3}}}}+\frac{1}{3\sqrt{\pi^{3}}}
\left(\sqrt{S_{4}^{3}}-\sqrt{S_{3}^{3}}\right)\nonumber\\
&-&\frac{C_{XY}^{5}}{12(C_{XY}+1)\pi^{1+\frac{1}{C_{XY}}}}\left(S_{4}^{1+\frac{1}{C_{XY}}}-S_{3}^{1+\frac{1}{C_{XY}}}\right)
-\frac{C_{XY}M\pi^{\frac{1}{2}-\frac{1}{C_{XY}}}}{3(C_{XY}-2)}\left(S_{4}^{\frac{1}{C_{XY}}-\frac{1}{2}}-S_{3}^{\frac{1}{C_{XY}}-\frac{1}{2}}\right)\nonumber\\
&+&\frac{C_{XY}Q^{2}\pi^{1-\frac{1}{C_{XY}}}}{12(C_{XY}-1)}\left(S_{4}^{\frac{1}{C_{XY}}-1}-S_{3}^{\frac{1}{C_{XY}}-1}\right).
\end{eqnarray}
For the new heat engine, the thermal efficiency can be written in
the following  form,
\begin{eqnarray}
\eta_{_{New}}=\frac{W}{Q_{H}}=\frac{(p_{1}-p_{4})(V_{2}-V_{1})}{Q_{H}},
\end{eqnarray}
which crucially depends on the MCCG parameters. In Fig. \ref{fig-eta} we can see typical behavior of the thermal efficiency for the various values of $B$ parameters and find that there is no important differences at large horizon radius.

\subsection{Rankine Cycle}
\begin{figure}[h!]
 \begin{center}$
 \begin{array}{cccc}
\includegraphics[width=70 mm]{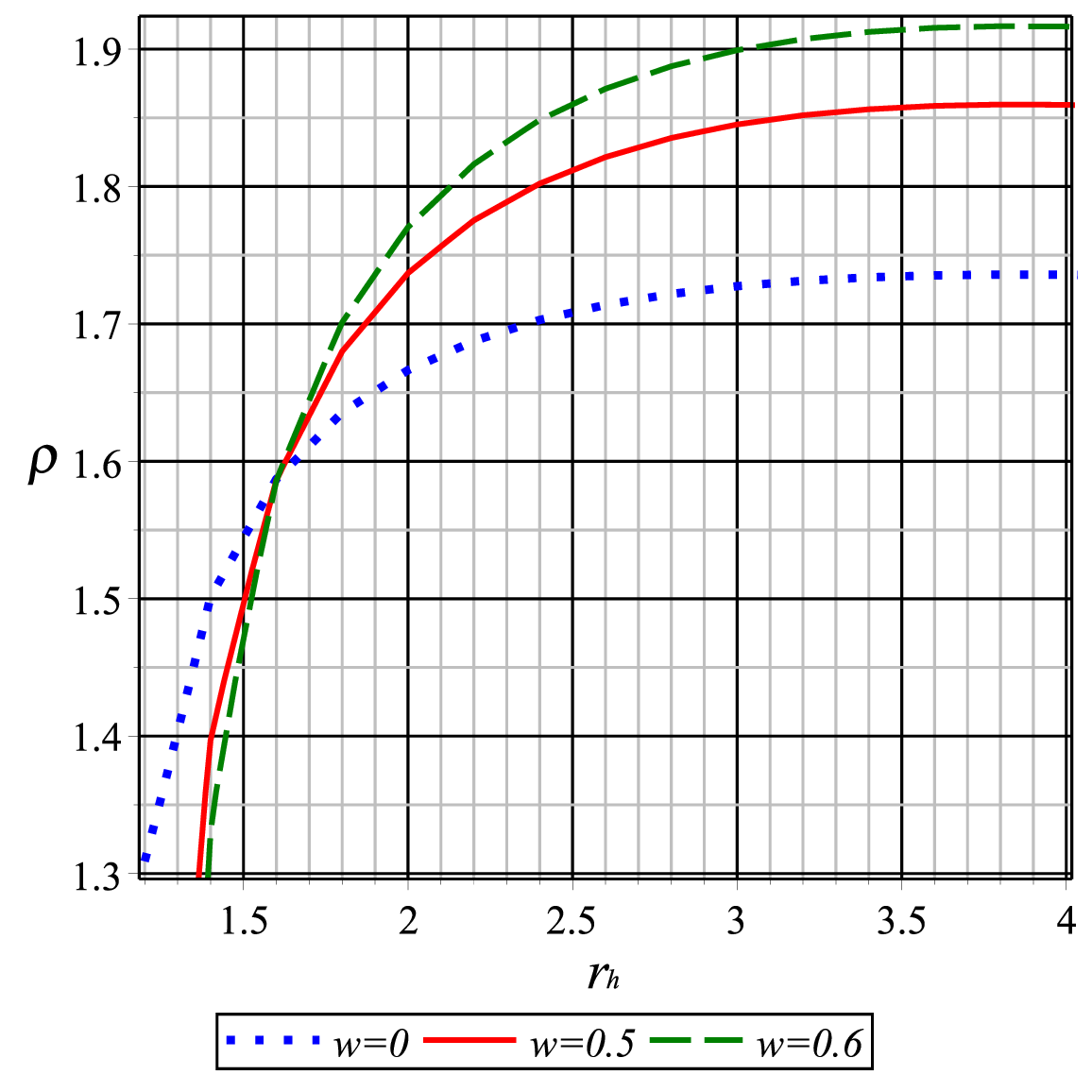}
 \end{array}$
 \end{center}
\caption{energy density in terms of horizon radius for $M=1$, $Q=1$, $C=1$, $A=1$, $l=1$ and $\alpha=0.5$.}
 \label{fig12}
\end{figure}

The Rankine cycle \cite{Wei1} is useful to predict the performance
of steam engine, which is an ideal thermodynamic cycle of a heat
engine that converts heat into mechanical work while undergoing
phase change. In each cycle, the working substance will undergo a
liquid-gas phase transition. The Rankine cycle for black hole heat
engine is shown by the Ref. \cite{Wei1}.\\
For the fixed pressure $p$ ($dp=0$), the first law of the black hole
system is $dH_{p}=TdS$ and so the enthalpy for constant pressure
is $H_{p}(S)=\int TdS$. So according to the Ref. \cite{Wei1}, the
efficiency for Rankine cycle can be written as,
\begin{equation}
\eta_{_{Ran}}=1-\frac{T_{1}(S_{3}-S_{1})}{H_{p_{_{2}}}(S_{3})-H_{p_{_{2}}}(S_{1})}
\end{equation}
where the subscripts 1 and 3 denote the lower process. Here $T_{1}$ and $H
_{p_{_{2}}}$ are given by,
\begin{equation}\label{T222}
T_{1}=\frac{1}{4\pi}\left[\frac{2\sqrt{S_{1}}}{l^{2}\sqrt{\pi}}+\frac{2\pi M}{S_{1}}-\frac{2\sqrt{\pi^{3}}Q^{2}}{\sqrt{S_{1}^{3}}}-\frac{C_{XY}}{3}e^{\frac{\sqrt{S_{1}}}{3\sqrt{\pi}}}
\exp{\left(\gamma_{1}(A\rho_{1}-\rho_{1}^{-\alpha}\left(C+(\rho_{1}^{1+\alpha}-C)^{-w} \right))\right)}\right],
\end{equation}
where
\begin{equation}\label{37-2}
\gamma_{1}=\frac{8\pi l^{2}(3+\sqrt{\frac{S_{1}}{\pi}})S_{1}^{2}}{27(\pi^{2}Q^{2}l^{2}+S_{1}^{2})},
\end{equation}
and
\begin{eqnarray}
H_{p_{_{2}}}(S_{i})&=&\frac{9\pi^{4}\left(Q^{2}+\frac{S_{i}^{2}}{\pi^{2}}(\frac{2}{3}-\frac{8}{9}\pi p_{2})\right)}{12\sqrt{\frac{S_{i}}{\pi}}}\nonumber\\
&-&\frac{S_{i}^{2}(S_{i}+\sqrt{\frac{S_{i}}{\pi}})\left(Q^{2}+\frac{S_{i}^{2}}{\pi^{2}}(1-\frac{16}{9}\pi p_{2})C_{XY}\exp{(\frac{\sqrt{S_{i}}}{3\sqrt{\pi}}+\gamma_{i}p_{2})}\right) }{12\sqrt{\frac{S_{i}}{\pi}}(S_{i}^{2}+\pi^{2}Q^{2})},
\end{eqnarray}
where $i=1,3$ and
\begin{equation}
p_{2}=A\rho_{2}-\rho_{2}^{-\alpha}\left(C+(\rho_{2}^{1+\alpha}-C)^{-w} \right)
\end{equation}
General behavior of the above variable are similar to that obtained in the section \ref(sec4).\\
Only remained thing is variation of density $\rho$ with the horizon radius $r_{h}$ and other model parameters like $w$. It is clear from the Fig. \ref{fig12} that energy density is increasing function of the horizon radius. Dotted blue line is corresponding to MCG black hole \cite{Debnath1}.

\section{Conclusion} \label{sec6}
In this work, we constructed the MCCG AdS black hole. For this purpose, we assumed a cosmological constant as a thermodynamic pressure and used MCCG equation of state. Then, we considered general spherically symmetric static ansatz and solved the field equations to derive the metric function. As a result, the MCCG equation of state emerges from the relevant thermodynamics, which is holographically dual to a charged AdS black hole. We studied the MCCG AdS black hole's horizon structure and  showed that it is possible to have both inner and outer horizons. Depending on the model parameters, there is also the possibility of having an extremal black hole or naked singularity. We discussed about the weak, strong and dominant energy conditions and found that all the energy conditions may be satisfied together by choosing appropriate radius. Then, we studied MCCG AdS black hole thermodynamics and calculated temperature, enthalpy, Gibbs and Helmholtz free energies. We discussed graphically about the behavior of thermodynamics variables with event horizon radius. We show that there is no any critical point. We discovered that depending on the horizon radius, the specific heat capacity can be positive or negative. Analyzing the specific heat capacity have shown that stable/unstable phase transition happen in this model. We confirmed that the case of uncharged black hole may completely unstable. This work is indeed extension of the Ref. \cite{Debnath1} by including cosmic parameter and the black hole charge. The cosmic parameter help us to remove some unstable regions, so that the model may completely stable. In Ref. \cite{Debnath1}, the approach utilized was power series expansion, however in this study, we used the separation variable method. Then, we computed the Joule-Thomson coefficient and discussed about cooling or heating processes. Finally, we considered MCCG AdS black hole as heat engine and calculated efficiencies. The efficiencies for Carnot cycle and Rankine cycle depend on the MCCG parameters $\alpha$, $w$, $A$ and $C$. For $\alpha=-1$ or $w=0$, we see that MCCG reduces to MCG and in this case the results are identical with \cite{Debnath1}. For MCCG, the efficiencies contain extra term $(\rho_{1}^{(1+\alpha)} - C)^{-w}$ rather than the MCG, so this term favours the reducing nature of efficiencies (rather than MCG) if we consider $S_1 > S_2$. So MCCG has vital role to determine the nature of efficiencies which depend on the model parameters.\\
In summary, we claim that the black hole described by
$$
ds^{2}=-fdt^{2}+\frac{dr^{2}}{f}+r^{2}d\Omega_{2}^{2},
$$
where
$$
f=\frac{r^{2}}{l^{2}}-\frac{2M}{r}+\frac{Q^{2}}{r^{2}}-Be^{\frac{r}{3}},
$$
is holographically dual of a MCCG.\\
As we stated before, this work is nothing but the extension of the recent work, Ref. \cite{Debnath1}, in order to include the cosmic parameter and black hole charge. The obtained black hole solution removes some instabilities of the Chaplygin gas model. However, it is seen that there are still some unstable regions depending on the model parameters of MCCG. Nevertheless, it is still possible to remove those unstable regions by extending the model. For example, one can consider bulk or shear viscosities. In the general case, one can consider extended Chaplygin gas \cite{e1, e2, e3, e4, e5, e6, e7, e8, e9, e10} and thus can use this paper as a guideline to have non-unstable regions. Also, it is interesting to obtain the effects of thermal fluctuations \cite{fl1, fl2, fl3, fl4, fl5, fl6, fl7, fl8} on the MCCG AdS black hole. All of them are on our to-do list for the foreseeable future.

\section*{ Acknowledgements}

The authors are grateful to the Editor and anonymous Reviewers for their
valuable comments and suggestions to improve the paper.

\end{document}